\documentclass[11pt]{article}
\pdfoutput=1
\usepackage{draft}
\usepackage{comment}
\usepackage[normalem]{ulem}
\usepackage{hyperref}
\hypersetup{
 colorlinks=true,
 urlcolor=blue,
 anchorcolor=blue,
 citecolor=blue,
 filecolor=blue,
 linkcolor=blue,
 menucolor=blue,
 pagecolor=blue,
 linktocpage=true,
}
\usepackage{graphicx,color,subfig}
\usepackage{cite}
\usepackage{empheq}
\usepackage[font=small,labelfont=bf]{caption} 

\usepackage{bbold}
\usepackage[usenames,dvipsnames,table]{xcolor}
\graphicspath{{./figures/}}
\usepackage{tikz}

\newcommand{\ZZ}{\mathbb{Z}}

\DeclareMathOperator{\arccosh}{arccosh}
\newcommand{\kernel}{\mathbb{K}}

\DeclareMathOperator{\vol}{vol}

\newcommand{\half}{{1\over 2}}

\DeclareFontFamily{OT1}{pzc}{}
\DeclareFontShape{OT1}{pzc}{m}{it}{<-> s * [1.10] pzcmi7t}{}
\DeclareMathAlphabet{\mathpzc}{OT1}{pzc}{m}{it}

\definecolor{vert}{rgb}{0.1367 0.543 0.1367}

\def\({\left(}
\def\){\right)}

\def\kernel{\mathbb{K}}

\newcommand{\mass}{\includegraphics[height=4.000mm, trim = 0mm 0mm 0mm 0mm, clip]{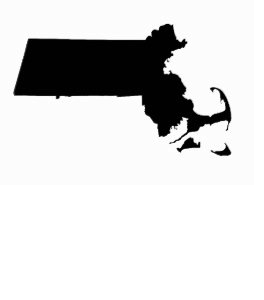}}
\newcommand{\nj}{\includegraphics[height=3.750mm, trim = 0mm 0mm 0mm 0mm, clip]{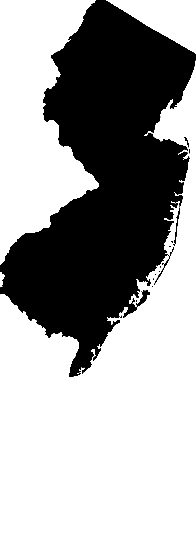}}

\newcommand{\beaver}{\includegraphics[height=2.000mm, trim = 0mm 0mm 0mm 0mm, clip]{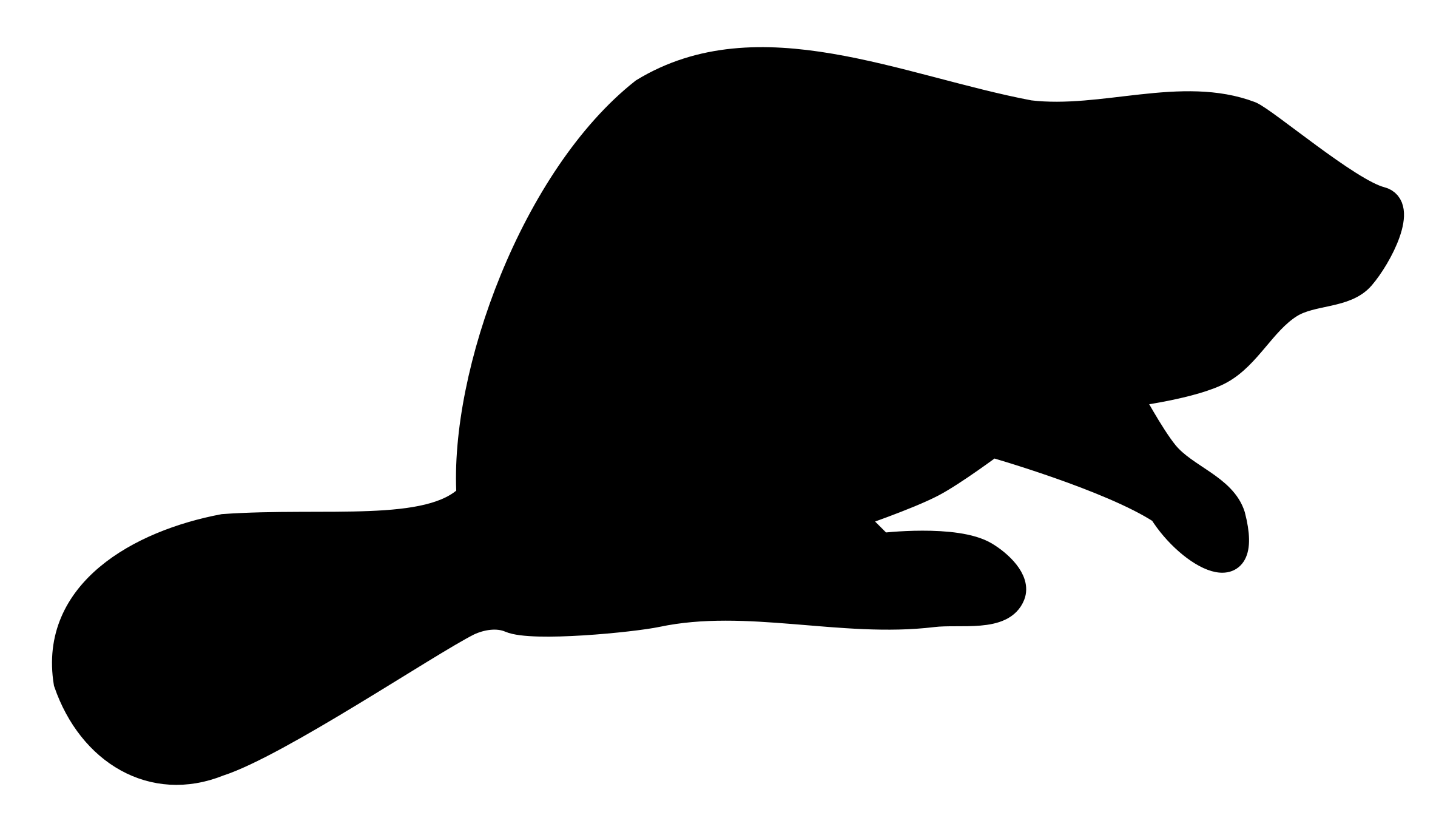}}

\begin{document}

\unitlength = .8mm

 \begin{titlepage}
 
		 \begin{flushright}
		\hfill{\tt PUPT-2612}
		\end{flushright} 
 
 \begin{center}

 \hfill \\
 \hfill \\

\title{Pure Gravity and Conical Defects}

 \author{Nathan Benjamin\nj, Scott Collier\mass, Alexander Maloney$^{\beaver}$}

 \address{
\nj Princeton Center for Theoretical Science, Princeton University, Princeton NJ 08544, USA\\
 \mass Jefferson Physical Laboratory, Harvard University, 
 Cambridge, MA 02138, USA
 \\
 $^{\beaver}$Department of Physics, McGill University,
 Montreal, QC H3A 2T8, Canada
 }

 \email{nathanb@princeton.edu, scollier@g.harvard.edu, alex.maloney@mcgill.ca}

 \end{center}

 \abstract{
We revisit the spectrum of pure quantum gravity in AdS$_3$.  The computation of the torus partition function will -- if computed using a gravitational path integral that includes only smooth saddle points -- lead to a density of states which is not physically sensible, as it has a negative degeneracy of states for some energies and spins.  We consider a minimal cure for this non-unitarity of the pure gravity partition function, which involves 
the inclusion of additional states below the black hole threshold.  
We propose a geometric interpretation for these extra states: they are conical defects with deficit angle $2\pi(1-1/N)$, where $N$ is a positive integer. 
That only integer values of $N$ should be included can be seen from a modular bootstrap argument, and leads us to propose a modest extension of the set of saddle-point configurations that contribute to the gravitational path integral: one should sum over orbifolds in addition to smooth manifolds. 
These orbifold states are below the black hole threshold and are regarded as massive particles in AdS, but they are not perturbative states: they are too heavy to form multi-particle bound states. We compute the one-loop determinant for gravitons in these orbifold backgrounds, which confirms that the orbifold states are Virasoro primaries. We compute the gravitational partition function including the sum over these orbifolds and find a finite, modular invariant result; this finiteness involves a delicate cancellation between the infinite tower of orbifold states and an infinite number of instantons associated with $PSL(2,{\mathbb Z})$ images.

 }

 \vfill

 \end{titlepage}

\eject

\begingroup

 \tableofcontents

\endgroup

\section{Introduction}\label{sec:intro}
 
One of the simplest possible theories of gravity is three dimensional general relativity  with negative cosmological constant, governed by the Einstein Hilbert action
\begin{equation}
  S = {1\over 16\pi G_N}\int d^3 x \sqrt{g}\(R+{2\over L^2}\).
\end{equation}
Although this theory has no local gravitons and all classical solutions are locally AdS$_3$ with radius $L$, it nevertheless has a rich structure including black hole solutions \cite{Banados:1992wn}.  
With asymptotically AdS boundary conditions, the theory possesses an infinite dimensional asymptotic Virasoro symmetry \cite{Brown:1986nw} with central charge 
\begin{equation}
  c = {3L\over 2 G_N} + \mathcal{O}(1).
\end{equation} 
This highly constrains the theory, and allows the use of the analytical technology of two dimensional conformal field theory to study foundational questions about quantum gravity in a potentially exactly solvable setting.

There is a sense in which every 2D CFT defines a theory of three-dimensional quantum gravity. However, for the bulk dual to resemble semiclassical Einstein gravity, the central charge must be large and the spectrum of light operators must be sufficiently sparse \cite{Witten:2007kt, Heemskerk:2009pn}. 
Under these assumptions, semiclassical Einstein gravity has many features that are universal. For example, the Cardy formula for the asymptotic density of states in 2D CFT \cite{Cardy:1986ie}, which reproduces the Bekenstein-Hawking entropy of black holes with an $AdS_3$ factor in the near-horizon geometry \cite{Strominger:1996sh,Strominger:1997eq}, has an extended regime of validity under these assumptions \cite{Hartman:2014oaa}, reflecting the validity of gravitational effective field theory in $AdS_3$.\footnote{The analysis in \cite{Hartman:2014oaa} actually only established the extended regime of validity of the Cardy formula for dimensions $\Delta \ge {c\over 6}$ in the case of large central charge and a sparse light spectrum; the gravitational effective field theory argument would suggest that it should hold for $\Delta \ge {c\over 12}(1+\epsilon)$, where $\epsilon$ is an order-one number.}

Indeed, the low energy spectrum of a pure theory of gravity -- namely one with only metric degrees of freedom -- includes only Virasoro descendants of the vacuum state.  These are interpreted as ``boundary gravitons," analogous to the edge modes of a quantum hall system, which correspond to perturbations of the AdS$_3$ metric which are not pure gauge only because of the existence of the asymptotic boundary. 
The BTZ black holes solutions correspond to states with mass above the Planck scale -- or dimensions $h$ and ${\bar h}$ greater than $c/24$ in the CFT language.  The natural expectation, therefore, is that pure quantum gravity should be a maximally sparse theory, whose spectrum includes only these boundary gravitons as well as heavy states with $h, {\bar h}>c/24$ which describe the microstates of these black holes \cite{Witten:2007kt}. Indeed, the question of how sparse the light spectrum can be consistent with the constraints of conformal symmetry, unitarity and locality of the dual CFT (or equivalently, consistency of quantum gravity in the bulk) is a topic of ongoing research without a definite conclusion (see e.g. \cite{Hellerman:2009bu,Friedan:2013cba,Collier:2016cls,Afkhami-Jeddi:2019zci,Hartman:2019pcd}).

In \cite{Maloney:2007ud,Keller:2014xba}, the authors attempt to compute the partition function of pure three-dimensional gravity via an explicit sum over known saddles in the Euclidean gravitational path integral. In CFT terms, the sum over gravitational instantons amounts to a sum over $PSL(2,\ZZ)$ images of the Virasoro vacuum character
\begin{equation}
  Z_{\rm MWK}(\tau,\bar\tau) = \sum_{ \gamma\in\mathbb Z \backslash PSL(2,\ZZ)}|\chi_{\rm vac}(\gamma\tau)|^2,
\end{equation}
a procedure generally known as a Poincar\'e series. The partition function thus obtained is a finite, modular-invariant function, but has the following unphysical features:
\begin{enumerate}[1)]
\item The spectrum of twists at fixed spin $j$ is continuous rather than discrete.
\item At $h=\bar{h}=\frac{c-1}{24}$, the spectrum has a degeneracy of $-6$. 
\item In \cite{Benjamin:2019stq}, it was pointed out that at any odd spin, for sufficiently low energy, the density of states is negative.
\end{enumerate}
While the first feature is perhaps in conflict with the conventional interpretation of pure 3d gravity as dual to a single quantum system,\footnote{This is more problematic than the continuous spectra encountered in familiar examples of non-compact CFTs such as Liouville theory and the dual of type IIB string theory on $AdS_3\times S^3\times M_4$ supported purely by NS-NS three-form fluxes \cite{Maldacena:2000hw,Eberhardt:2019qcl} because here the vacuum is a normalizable state.} the last two features signal non-unitarity are more problematic.
In this paper we will describe a simple modification of \cite{Maloney:2007ud} which is positive everywhere.  We will moreover provide a natural physical interpretation of the new contributions that render the spectrum unitary. 

Several other proposed resolutions of problems (2) and (3) have been considered in the literature.  In \cite{Keller:2014xba}, it was noted that problem (2) can be cured with the addition of six $c=1$ free boson partition functions to the spectrum.\footnote{An entirely different  --  although highly non-minimal -- proposal to cure these problems was considered in \cite{Maloney:2007ud}, which is to sum separately over two sets of $PSL(2, {\mathbb Z})$ instantons, which act on left- and right-moving sectors separately.  The bulk interpretation of this is not entirely clear, however, as these lead to saddles which -- although smooth as $SL(2,{\mathbb R})\times SL(2,{\mathbb R})$ Chern-Simons connections -- are both complex and singular from the metric point of view.}
In \cite{Benjamin:2019stq}, problem (3) was cured with the addition of operators with sufficiently low twist (in particular with $\text{min}(h, \bar h) \leq \frac{c-1}{32}$). In this paper we will calculate the {\it minimal} operator content needed to cure (3). In particular, we will attempt to maximize the dimension and minimize the degeneracy of each additional operator we add. It turns out that at large $c$ the minimal prescription involves the addition of operators with conformal dimensions\footnote{We use the standard notation where the left- and right-moving conformal dimensions are related to the conformal dimension and spin by $\Delta=h+{\bar h}$ and $J=h-{\bar h}$.}
\be\label{frodo}
h = \bar h = \frac{c}{24}\(1-\frac1{N^2}\) + \mathcal O(c^0), ~~~~N = 2, 3, \ldots.
\ee
We will discover that these states have a beautiful interpretation in terms of AdS$_3$ gravity.
In particular, equation (\ref{frodo}) is precisely the correct conformal dimension to be interpreted as a conical defect geometry which is a quotient of AdS$_3$ by $\mathbb Z_N$.  From the bulk point of view, this is a massive particle which produces a deficit angle of $2\pi(1-\frac 1N)$.  We will see that these orbifolds, with positive integer values of $N$, can be consistently included in the Euclidean path integral of three dimensional general relativity.\footnote{In \cite{Krasnov:2000ia, Krasnov:2001ui, Krasnov:2002rn}, it has been suggested that point particle states may be required for the consistency of quantum gravity in AdS$_3$.}

We will then proceed to construct the full spectrum of the theory by including these as saddle points in the sum over geometries that computes the torus partition function.   We will argue that the gravitational path integral is one-loop exact (just as for the smooth saddle points which contribute to the torus path integral in \cite{Maloney:2007ud}).  We will compute the resulting spectrum that arises from the sum over geometries.   
We show that the inclusion of a finite number of orbifolds in the sum leads to a density of states that is finite and positive everywhere except at the threshold value of $h={\bar h} = {c-1\over 24}$.  This negativity at threshold was encountered previously in the literature \cite{Keller:2014xba} and is easily cured with the addition of a free boson partition function. The result is a spectrum that is finite and continuous (in the regime $h,{\bar h}>{c-1\over 24}$), but is everywhere positive for sufficiently large central charge.  The fact that the spectrum is continuous reflects that it cannot be dual to a single, unitary, compact CFT, but it is possible that it can be interpreted as some kind of ``averaged" partition function, as is known to occur in two dimensions \cite{Saad:2019lba, Stanford:2019vob}.  We also consider the more general case where the infinite family of orbifolds with $N=2,3,\dots$ are included in the sum over geometries.
At first sight it might appear that the result must diverge, as the states (\ref{frodo}) accumulate at $h={\bar h}={c-1\over24}$, resulting in an infinite number of states at this critical value of the dimension.  We will, however, see that this is precisely cancelled by a divergence (coming with the opposite sign) that arises from the sum over modular images.  In this way the contributions of two infinite families of instantons -- the $PSL(2,{\mathbb Z})$ saddles and the orbifolds -- combine to give a candidate pure gravity partition function that is finite, although the resulting density of states does not appear to be well-defined. Another possibility is that one must truncate the sum over orbifolds in a central charge dependent way so that the full set of states is only included at infinite $c$; a natural choice would be at $N\sim \mathcal{O}(c^{1/2})$, c.f. (\ref{frodo}).

This paper is organized as follows. In section \ref{sec:negativity}, we review the partition function of \cite{Maloney:2007ud, Keller:2014xba} and rewrite the spectrum as a sum over modular crossing kernels. We also write down a minimal addition of operators to render the spectrum positive everywhere. In section \ref{sec:orbifolds} we provide a physical interpretation of the extra states we add as singular orbifolds that contribute to the path integral of pure gravity in AdS$_3$.  We will argue that this path integral must be one-loop exact.  In section \ref{sec:oneLoop}, we perform a direct one-loop computation to see the $1/c$ of effect of these orbifold geometries to the path integral. Finally in section \ref{sec:discuss} we conclude and discuss possible interpretations of our partition function.

\section{Non-unitarity of the pure gravity partition function}\label{sec:negativity}

We will start by reviewing the construction of the so-called MWK partition function of pure AdS${}_3$ gravity, which was first discussed in \cite{Maloney:2007ud} and studied further in \cite{Keller:2014xba}.  This partition function is obtained by computing a sum over Euclidean geometries with torus boundary, including in the sum all metrics that are continuously connected to a smooth saddle point. This sum includes thermal $AdS_3$, the Euclidean BTZ black hole and the so-called $PSL(2,\mathbb{Z})$ family of black holes. The $PSL(2,\mathbb{Z})$ family of black holes can be thought of as generalizations of the BTZ black hole with different linear combinations of the boundary cycles taken to be contractible. At the level of the partition function of the boundary theory, this amounts to a sum over $PSL(2,\mathbb{Z})$ images of the Virasoro vacuum character. So while the construction of the boundary partition function via a Poincar\'e series may seem ad hoc from the point of view of the CFT, it has a natural physical interpretation as the sum over saddles in the Euclidean gravitational path integral. As discussed in \cite{Benjamin:2019stq}, the microcanonical density of states one obtains from this procedure becomes negative for odd spin primaries of high spin and low twist.  One way to cure this negativity is by the addition of states with twist at or below $\frac{c-1}{16}$ to the seed spectrum. In this section we consider the ``minimal" set of operators that can be added to render the spectrum unitary. Although in this paper we focus on the large $c$ limit, we note that many of the results in this section only require $c>1$.

\subsection{Review of the MWK spectrum}
\label{sec:ReviewMWK}

The MWK partition function is obtained by taking the Virasoro vacuum character -- which is interpreted as the partition function of a gas of boundary gravitons in thermal AdS$_3$ -- and summing over $PSL(2,\mathbb Z)$ images. Although the sum diverges, \cite{Maloney:2007ud, Keller:2014xba} used zeta-function regularization to give a finite, modular-invariant answer with a continuous density of states at each integer spin. We will give a slightly different derivation of this result, where rather than working with the partition function we will work directly with the density of states.  Thus instead of computing the partition function as a sum over $PSL(2,\mathbb Z)$ images of the vacuum character, we compute directly the contribution to the density of states coming from one of the $PSL(2,\mathbb{Z})$ images of the vacuum state in terms of a modular crossing kernel, and then sum over $PSL(2,\mathbb{Z})$.   

The modular crossing kernel $\mathbb{K}^{(\gamma)}$ for an element $\gamma$ of $PSL(2,\mathbb{Z})$ expresses the modular transform of a Virasoro character $\chi_h(\gamma\tau)$ in a basis of untransformed characters: 
\begin{equation}
  \chi_{h}(\gamma\tau) = \int_{\frac{c-1}{24}}^\infty dh'\, \mathbb{K}^{(\gamma)}_{h'h}\chi_{h'}(\tau),
\end{equation}
where $\gamma\tau = {a\tau+b\over s\tau +d}$ for ${\small\begin{pmatrix}a & b\\ s& d\end{pmatrix}}\in PSL(2,\mathbb{Z})$.\footnote{We denote the element of $PSL(2,\mathbb Z)$ as $\small\begin{pmatrix} a & b \\ s & d \end{pmatrix}$ to avoid confusion with the central charge $c$.} 
The Virasoro character is
\begin{equation}
  \chi_h(\tau) = {q^{h-\frac{c-1}{24}}\over\eta(\tau)},
\end{equation}
where $q = e^{2\pi i\tau}$ and the explicit expression for the modular crossing kernel for $s>0$ is\footnote{We thank Henry Maxfield for bringing the explicit expression for the general $PSL(2,\ZZ)$ modular crossing kernel to our attention. See also appendix D of \cite{Benjamin:2019stq} for the kernels with $b=0$.}
\begin{equation}\label{eq:sl2zKernel}
  \mathbb{K}^{(\gamma)}_{h'h} = {\epsilon(\gamma)}\sqrt{\frac2s}e^{{2\pi i\over s}(a(h-\frac{c-1}{24})+d(h'-\frac{c-1}{24}))}{\cos\({4\pi \over s}\sqrt{(h-\frac{c-1}{24})(h'-\frac{c-1}{24})}\)\over\sqrt{h'-\frac{c-1}{24}}}.
\end{equation}
Here $\epsilon(\gamma)$ is a $h,h'$-independent phase that will be unimportant since we will be considering products of holomorphic and antiholomorphic Virasoro characters. 

The MWK partition function of \cite{Maloney:2007ud} and \cite{Keller:2014xba} is obtained by summing over $PSL(2,\mathbb{Z})$ images of the Virasoro vacuum character
\begin{equation}\label{eq:vacuumCharacter}
  \chi_{\rm vac}(\tau,\bar \tau) = (\chi_0(\tau)-\chi_1(\tau))(\overline{\chi}_0(\bar\tau)-\overline{\chi}_1(\bar\tau)) = {\left|q^{-\frac{c-1}{24}}(1-q)\right|^2\over|\eta(\tau)|^2}.
\end{equation}
 Aside from the vacuum Verma module itself, the spectrum has support only on primaries above the BTZ threshold, $\text{min}(h,\bar h) \ge \frac{c-1}{24}$.  Thus one can write the density of states of the MWK partition function simply as a $PSL(2,\mathbb{Z})$ sum over the modular crossing kernels, as: 
\begin{equation}
  \rho^{\rm MWK}(h,\bar h) = \sum_{\gamma\in \mathbb Z \backslash PSL(2,\ZZ)}\left[\kernel^{(\gamma)}_{h0}\overline{\kernel}^{(\gamma)}_{\bar h 0}-\kernel^{(\gamma)}_{h0}\overline{\kernel}^{(\gamma)}_{\bar h 1} -\kernel^{(\gamma)}_{h1}\overline{\kernel}^{(\gamma)}_{\bar h 0} + \kernel^{(\gamma)}_{h1}\overline{\kernel}^{(\gamma)}_{\bar h 1}\right].
\label{eq:sumgamma}
\end{equation}
In appendix \ref{app:regs} we show that the density of states (\ref{eq:sumgamma}) can easily be decomposed into sectors with integer spin. We then arrive at the following density of states (see also (4.11) of \cite{Benjamin:2019stq})\footnote{In what follows, we will often make use of parity-invariance and assume without loss of generality that $j\geq0$.}
\begin{equation}\begin{aligned}\label{eq:MWKDensity2}
 \rho^{\rm MWK}_{j}(t)= {2\over \sqrt{t(t+j)}}\sum_{s=1}^\infty{1\over s}&\bigg[S(j,0;s)\cosh\({{4\pi \over s}\sqrt{\frac{c-1}{24} (t+j)}}\)\cosh\({{4\pi \over s}\sqrt{\frac{c-1}{24} t}}\)  
  \\ &-S(j,-1;s)\cosh\({{4\pi\over s}\sqrt{\frac{c-1}{24} (t+j)}}\)\cosh\({{4\pi\over s}\sqrt{\frac{c-25}{24} t}}\)
  \\& -S(j,1;s)\cosh\({{4\pi \over s}\sqrt{\frac{c-25}{24}(t+j)}}\)\cosh\({{4\pi \over s}\sqrt{\frac{c-1}{24} t}}\) 
  \\&+S(j,0;s)\cosh\({{4\pi\over s}\sqrt{\frac{c-25}{24} (t+j)}}\)\cosh\({{4\pi\over s} \sqrt{\frac{c-25}{24} t}}\)\bigg],
\end{aligned}\end{equation}
where $S(j, J; s)$ is a Kloosterman sum
\be
S(j,J;s) = \sum_{0\le d < s,~ \text{gcd}(s,d) = 1}\exp\(2\pi i \frac{dj+(d^{-1})_s J}{s}\),
\label{eq:kloos}
\ee
and we have written the density of states in terms of what we will refer to as the \emph{reduced twist}
\begin{equation}
  t \equiv \text{min}(h,\bar h)-\frac{c-1}{24}.
\end{equation}
This density of states (\ref{eq:MWKDensity2}) is only defined for $t\ge 0$, i.e. it only has support on twists above the BTZ threshold.

\begin{figure}[h!]
  \centering\includegraphics[width=7cm]{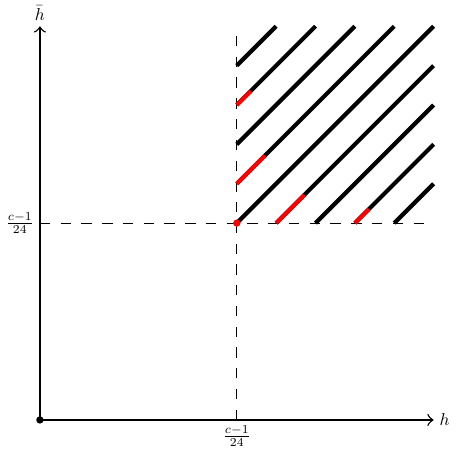}
  \caption{The spectrum of the MWK partition function. We start with the vacuum state, at $h=\bar h = 0$ and get a continuous spectra at all integer spins, with $h, \bar{h} \geq \frac{c-1}{24}$. The spectrum is not unitarity; the red regions represent the parts of the spectrum where the density of states is negative.}
  \label{fig:boat}
\end{figure}

The sum over $PSL(2,{\mathbb Z})$ described above is known as a Poincar\'e series, and the procedure described above is an algorithm that takes a seed state and computes a modular invariant spectrum that is otherwise only supported above the BTZ threshold $t\ge0$ (except for the seed state, of course).  As we will see, however, unitarity of the resulting spectrum is not guaranteed. See figure \ref{fig:boat} for a sketch of the spectrum. 

More generally, we can consider the contribution to the density of states of the $PSL(2,\mathbb{Z})$ sum of the contribution of an arbitrary light seed operator with reduced twist $T<0$ and spin $J$. The above procedure leads to the following density of states in the spin-$j$ sector for $jJ\geq0$ 
\begin{equation}\begin{aligned}\label{eq:generalPoincare}
  \rho_j^{(T,J)}(t) =& {2\over\sqrt{t(t+|j|)}}\sum_{s=1}^\infty{1\over s}S(j,J;s)\cosh\({4\pi\over s}\sqrt{-(T+|J|)(t+|j|)}\)\cosh\({4\pi \over s}\sqrt{-T t}\).
\end{aligned}\end{equation}
On the other hand, if $jJ\leq0$ we have
\begin{equation}\label{eq:generalPoincare2}
   \rho_j^{(T,J)}(t) = {2\over\sqrt{t(t+|j|)}}\sum_{s=1}^\infty{1\over s}S(j,J;s)\cosh\({4\pi\over s}\sqrt{-(T+|J|)t}\)\cosh\({4\pi \over s}\sqrt{-T(t+|j|)}\).
\end{equation}

The sums (\ref{eq:MWKDensity2}), (\ref{eq:generalPoincare}) and (\ref{eq:generalPoincare2}) are divergent, as the summands behave like $\mathcal O(s^{-1})$ at large $s$. In appendix \ref{app:regs} we  describe a regularization procedure that renders them finite, analogous to the zeta function regularization of \cite{Maloney:2007ud,Keller:2014xba}. We also show that this regularized sum of modular kernels is precisely the MWK spectrum \cite{Maloney:2007ud, Keller:2014xba}.\footnote{In the recent paper \cite{Alday:2019vdr}, a different prescription based on a technique by Rademacher was described that does \emph{not} have this divergence in $s$.  One of the $\cosh$ factors is replaced with a $\sinh$, leading to a different finite, modular-invariant partition function with the same censored (twist below the black hole threshold) part of the spectrum. In the case of a scalar seed, the Rademacher expansion can be interpreted as a Poincar\'e series where the seed is taken to be the original character plus a continuous density of scalar characters above the black hole threshold. It is not obvious to us whether the Rademacher expansion with a non-scalar seed can also be interpreted as a Poincar\'e series. We thank L. F. Alday for discussions on this point.} The regularization procedure produces a negativity at threshold in the scalar sector of the theory at $h=\bar{h} = \frac{c-1}{24}$ discussed in \cite{Maloney:2007ud, Keller:2014xba}:
\be
\rho_{j=0}^{\text{MWK}}(t) = -6\delta(t) + (\text{a continuous function of }t),
\ee
which we re-derive in appendix \ref{app:regs} using the modular kernels.

In \cite{Keller:2014xba} an ad hoc prescription was offered to cure this negativity, namely via the addition of six copies of the compact boson partition function to the pure gravity partition function. Besides this discrete negative degeneracy in the scalar sector, the MWK density of states in each spin sector is a smooth function of the twist supported only above the BTZ threshold $t>0$. 

The derivation described above allows us to easily understand the negativity in the near-extremal, large-spin regime recently discussed in \cite{Benjamin:2019stq}. 
From (\ref{eq:MWKDensity2}), we see that there is a simple expression for the near-extremal ($t\to 0$), large-spin  ($j\to\infty$) limit of the MWK density of states
\begin{align}
\rho^{\rm MWK}_{j}(t) \sim &{8\pi^2\over\sqrt{j}}\sqrt{t}\exp\(4\pi\sqrt{\({c-1\over 24}\)j}\)
+\sum_{s=2}^\infty \Bigg[\frac{S(j, 0; s) - S(j,-1;s)}{s\sqrt{jt}} \exp\(\frac{4\pi}{s}\sqrt{\(\frac{c-1}{24}\) j}\) \nonumber\\&
+ \frac{S(j, 0; s) - S(j,1;s)}{s\sqrt{jt}} \exp\(\frac{4\pi}{s}\sqrt{\(\frac{c-25}{24}\) j}\)\Bigg].
\label{eq:blah}
\end{align}
The first term in (\ref{eq:blah}) exhibits the usual square-root edge in the reduced twist near extremality at large spin, as noted in \cite{Benjamin:2019stq,Maxfield:2019hdt}. Note that when the twist is sufficiently small, in particular for $t\lesssim {1\over 8\pi^2}e^{-2\pi\sqrt{\({c-1\over 24}\)j}}$, the second and third terms will dominate over the first.
We see that if the coefficient $S(j,0;s) - S(j,-1;s)$ is negative, then there may be a regime where the density of states is negative.\footnote{The asymmetry between $S(j,0;s) - S(j,-1;s)$ and $S(j,0;s) - S(j,1;s)$ in this expression arises because we take $j>0$. If we instead took $j<0$, the role of the two terms would be exchanged.}

To expand on this point, we compare the above expression to the contribution to the partition function of a Poincar\'e series starting with a seed of reduced twist $T<0$ and spin $J=0$.   This contributes to the density of states as
\be
\rho^{(T,J=0)}_j(t) \sim \sum_s \frac{S(j,0;s)}{s\sqrt{jt}}\exp{\(\frac{4\pi}{s}\sqrt{-Tj}\)}.
\label{eq:woof}
\ee
in the limit under consideration. The generalization to seed states with $J\ne0$ is straightforward, using  (\ref{eq:generalPoincare}) and (\ref{eq:generalPoincare2}).

We conclude that if $S(j,0;n)-S(j,-1;n)$ is negative for some $j,n$, it contributes a negativity to the partition function of the same magnitude as a state with reduced twist $T_n$ such that
\be\label{eq:orbifoldTwist}
T_n = -\frac{c-1}{24 n^2}.
\ee
Similarly, if $S(j,0;n)-S(j,1;n)$ is negative, for some $j,n$, it contributes a negativity to the partition function of the same magnitude as a state with twist $\widetilde T_n$ such that
\be
4\pi\sqrt{-\widetilde T_nj} = \frac{4\pi}n\sqrt{\(\frac{c-25}{24}\) j},
\ee
which means
\be\label{eq:orbifoldTildeTwist}
\widetilde T_n =  -\frac{c-25}{24 n^2} = T_n + \frac{1}{n^2}.
\ee
 A crucial point, and one which will be described in more detail in section \ref{sec:orbifolds}, is that -- to leading order in $c$ -- the reduced twists $T_n$ and $\widetilde T_n$ precisely correspond to the orbifold geometries discussed in section 2 of \cite{Maloney:2007ud}. 

\subsection{Minimalist spectrum}
\label{sec:minimalist}

In this section we will describe the minimal number of states needed to add to the MWK spectrum to render it positive. For simplicity we will only add scalar seeds. We will (first) maximize the dimension and (second) minimize the degeneracy of every additional scalar operator we add. Remarkably at sufficiently large central charge, it appears that adding a \emph{finite} number of operators is enough to render the entire spectrum positive above threshold. In the notation of the (\ref{eq:orbifoldTwist}), (\ref{eq:orbifoldTildeTwist}), we only need to add operators up at $T_2, \widetilde T_2, T_3, T_4, T_5, \widetilde T_5, T_6$, and $T_7$. In appendix \ref{app:howmany} we will explicitly show this calculation; here we will simply write down the result.\footnote{This table of degeneracies has been independently discovered by L. F. Alday and J.-B. Bae \cite{Alday:2019vdr}.}
\begin{table}
\begin{center}
  \begin{tabular}{ | c |  c | c |}
    \hline
    State & Conformal weights &  Minimal ``degeneracy''  \\ \hline
    $T_2$ & $h=\bar h =\frac{c-1}{32}$ & 1 \\ 
    $\widetilde T_2$ & $h=\bar h =\frac{c+7}{32}$ & 1 \\ 
    $T_3$ & $h=\bar h =\frac{c-1}{27}$ & 1 \\ 
    $T_4$ & $h=\bar h =\frac{5(c-1)}{128}$ & 1 \\
    $T_5$ & $h=\bar h =\frac{c-1}{25}$ & $\frac{5+\sqrt 5}{10}$ \\
   $\widetilde T_5$ & $h=\bar h =\frac{c}{25}$ & $\frac{5-\sqrt 5}{10}$ \\
    $T_6$ & $h=\bar h =\frac{35(c-1)}{864}$ & 1 \\
    $T_7$ & $h=\bar h =\frac{2(c-1)}{49}$ & $\frac{3+4\sin{\(\frac{3\pi}{14}\)}}{7} $ \\
    \hline
  \end{tabular}
\end{center}
\caption{Minimum number of scalars we need to add to render the MWK partition function finite.}
\label{tab:beavers}
\end{table}

In Table \ref{tab:beavers}, we write down the degeneracy and weight of every seed scalar operator needed. To each term in Table \ref{tab:beavers}, we perform a Poincar\'e sum as in (\ref{eq:generalPoincare}) and add it to the MWK partition function. Finally we add the partition function of $12 + \frac{3+4\sin{\(\frac{3\pi}{14}\)}}{7}$ self-dual free boson partition functions (to cancel the delta function negativities at threshold in the scalar sector). In appendix \ref{app:howmany} we will show that if the central charge is sufficiently large, this spectrum is positive everywhere; numerically we find that the cutoff is $c \gtrsim 3237.7$.

Of course the degeneracies in Table \ref{tab:beavers} are not integers, so this spectrum cannot be interpreted as a single compact, unitary CFT. On the other hand, the spectrum above the BTZ threshold is already continuous so we have already abandoned compactness.\footnote{By non-compact we simply mean that the spectrum is not given by a discrete sum of delta functions. However, the spectrum still has a normalizable $PSL(2,\mathbb{C})$ invariant vacuum, unlike other non-compact CFTs such as Liouville theory.} We will comment more on possible interpretations in section \ref{sec:discuss}.\footnote{If we demand that the degeneracies of all the $T_n$, $\widetilde T_n$ states are integers, we can have $1$ state at $T_2$, $\widetilde T_2$, $T_3$, $T_4$, and $T_5$, as well as 11 free boson partition functions, and the resulting partition function will be positive for $c > 1465.4$.}

\section{A geometric interpretation for the missing states}\label{sec:orbifolds}

In this section we will discuss how these orbifold states can be included directly in the path integral formulation of pure gravity in AdS$_3$.  We will follow the construction of MWK, who computed the partition function of pure gravity by enumerating the known saddle points in the gravitational path integral, along with an appropriate series of quantum corrections:
\begin{equation}\label{smaug}
Z(\tau,{\bar \tau})\equiv \int_{\partial M=T^2} Dg~ e^{-S} = \sum_{g_0} e^{-c S(g_0) +S^{(1)}(g_0)+\dots}
\end{equation}
We are computing here a sum over geometries whose boundary is a torus with modular parameter $\tau$, so that the the partition function can be used to extract the density of states $\rho(h, {\bar h})$.  Here $S(g_0)$ denotes the classical action of a classical saddle $g_0$, $S^{(1)}$ the one-loop correction to this action, and $\dots$ the infinite series of subleading perturbative corrections. The basic conjecture of MWK is that if the complete set of classical solutions $g_0$, along with the infinite series of perturbative corrections $\dots$ are included in the sum, then the saddle point approximation (\ref{smaug}) is exact.  This amounts to an assumption about the nature of the path integral of quantum gravity -- that it includes only metrics which are continuously connected to saddle points.    This is a familiar feature of many quantum field theories (especially Chern-Simons and topological field theories) whose partition function can be computed exactly, but should nevertheless be regarded as an assumption.

In the original MWK computation, the classical saddle points which contribute to this sum were taken to be the smooth Euclidean geometries which solve Einstein's equation (so are locally Euclidean AdS$_3$) and have torus boundary.  These geometries were completely classified, and can be interpreted physically as thermal $AdS_3$ and the Euclidean continuations of the $PSL(2,\mathbb{Z})$ family of BTZ black holes \cite{Maldacena:1998bw,Dijkgraaf:2000fq}.  These latter saddle points are, in Euclidean signature, related to the thermal AdS saddle by $PSL(2,{\mathbb Z})$ modular transformations. The perturbative corrections to these saddles can be computed by noting that in pure gravity the thermal AdS partition function will, at the perturbative level,  receive contributions only from multi-graviton states which are entirely fixed by Virasoro symmetry.   The result is that the perturbative partition function is one-loop exact, and one can perform the sum over geometries explicitly.  However, as reviewed in the previous section, the resulting spectrum is not unitary, as the density of states is negative in the near-extremal, large-spin regime. 

Here, we will advocate a small departure from the above construction:  we will relax the assumption that the saddle point geometries $g_0$ must be manifolds.  We will instead consider the inclusion of orbifold geometries, which take the form of quotients of AdS$_3$.   We will see that the orbifolds with torus boundary can be completely classified, just as in the previous discussion.  The result will be that the sum over geometries now includes additional gravitational instantons corresponding to ${\mathbb Z}_N$ quotients of thermal AdS, along with their images under $PSL(2,{\mathbb Z})$ modular transformations. We will also see that the perturbative corrections to these classical saddle points are one-loop exact.  The resulting gravity partition function can then be computed, and has a completely positive density of states in the dangerous near-extremal, large-spin regime.

In \cite{Maloney:2007ud} the classical solutions to the equations of motion of three dimensional gravity with negative cosmological constant and torus boundary were classified. Here we briefly review that classification, before generalizing to include orbifolds.  Any solution $M$ is a quotient $M = \widetilde{AdS}_3/\Gamma$, where $\Gamma$ is a subgroup of $SO(1,3)$ that acts discretely on $\widetilde{AdS}_3\subset AdS_3$.\footnote{Here we slightly abuse notation by using $AdS_3$ to denote Euclidean Anti-de Sitter space (i.e. ${\mathbb H}_3$) rather than Lorentzian AdS.} The conformal boundary of $M$, topologically a torus, can then also be written as a quotient
\begin{equation}
  \Sigma = U/\Gamma,
\end{equation}
where $U\subset \mathbb{CP}^1$ is the conformal boundary of $\widetilde{AdS}_3$, which is acted on by $SO(1,3)=SL(2,{\mathbb C})$ in the usual way by fractional linear Mobius transformations. One can then show \cite{Maloney:2007ud} that the fundamental group of $U$ must be
\begin{equation}
  \pi_1(U) \cong \ZZ.
\end{equation} 
In this case $U$ is topologically a cylinder and $\Gamma$ is a discrete subgroup of the group of diagonal matrices. The semiclassical solutions then split into two qualitatively different classes, which we briefly describe below. 

In the first class $\Gamma \cong \ZZ$, which was the case primarily considered in \cite{Maloney:2007ud,Keller:2014xba}. This class of solutions corresponds to thermal $AdS_3$ and the $PSL(2,\ZZ)$ family of BTZ black holes \cite{Maldacena:1998bw}. In this case $\Gamma$ is (up to an overall conjugation) generated by the diagonal matrix
\begin{equation}\label{eq:GammaGeneratorX}
  X = \begin{pmatrix} q & 0 \\ 0 & q^{-1} \end{pmatrix},
\end{equation}
where $q = e^{2\pi i\tau}$ is the modulus of the boundary torus $\Sigma$, with $|q|<1$. Of course, $\tau$ only defines inequivalent boundary Riemann surfaces up to $PSL(2,\ZZ)$ transformations
\begin{equation}
  \tau \to {a\tau+b\over s \tau+d},~\begin{pmatrix} a & b \\ s & d\end{pmatrix}\in PSL(2,\ZZ).
\end{equation}
Furthermore, as pointed out by \cite{Maloney:2007ud}, the family of three-manifolds solving the equations of motion are actually only labelled by a pair of coprime integers $(s,d)$; $(a,b)$ are only uniquely fixed by the $PSL(2,\ZZ)$ condition $ad-bs = 1$ up to integer shifts of $(a,b)$ by $(s,d)$ since the resulting $q$ is invariant under such shifts. Following \cite{Maloney:2007ud}, we refer to such three-manifolds as $\mathcal{M}_{s,d}$. They can be thought of as generalizations of the BTZ black hole with different linear combinations of the boundary cycles taken to be contractible.

It will be instructive to briefly describe the geometry of $\mathcal{M}_{0,1}$ in more detail. The Euclidean $AdS_3$ metric is given by
\begin{equation}\label{eq:euclideanMetric}
	ds^2 = dr^2 + \cosh^2 r\, dT^2 + \sinh^2 r\,d\phi^2,
\end{equation}
where $r\in[0,\infty)$, $T\in(-\infty,\infty)$ and $\phi \in [0,2\pi)$, with the conformal boundary at $r=\infty$. To quotient by the action of $X$ as in (\ref{eq:GammaGeneratorX}), the geometry is cut at $T = 0$ and $T = 2\pi\tau_2$ and glued upon making a rotation by the angle $2\pi \tau_1$. The resulting geometry is what we refer to as thermal $AdS_3$. In section \ref{sec:oneLoop} we will find it convenient to make use of the alternate coordinates $\rho = e^T$ and $\csc^2\theta = \cosh^2 r$, so that $\rho \in [1,e^{2\pi\tau_2}]$ and $\theta\in(0,{\pi\over 2}]$.

In the other class of solutions $\Gamma\cong \ZZ\times \ZZ_N$, for $N$ an integer greater than or equal to 2. In this case $\Gamma$ is generated by the following matrix in addition to (\ref{eq:GammaGeneratorX})
\begin{equation}\label{eq:GammaGeneratorY}
  Y = \begin{pmatrix} e^{2\pi i \over N} & 0 \\ 0 & e^{-{2\pi i \over N}} \end{pmatrix}.
\end{equation}

\begin{figure}[h!]
\centering
{
\subfloat{\includegraphics[width=.3\textwidth]{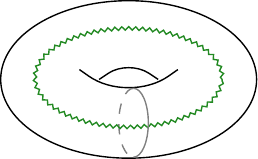}}\quad\quad\quad
\subfloat{\includegraphics[width=.18\textwidth]{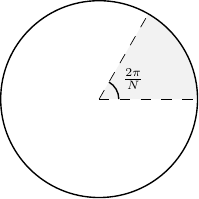}}
}
\caption{\textbf{Left: }The Euclidean geometry is asymptotically $AdS_3$ but has an orbifold singularity in the interior (shown in green). \textbf{Right: } The orbifold singularity corresponds to a conical defect of $2\pi(1-1/N)$.}\label{fig:singularBulk}
\end{figure}

This class of solutions includes geometries that are (singular) $\mathbb{Z}_N$ orbifolds of thermal $AdS_3$ and of the $PSL(2,\mathbb{Z})$ family of black holes. The possibility of including these in the gravitational path integral was briefly considered in \cite{Maloney:2007ud}, before being discarded in favor of the ``minimal" strategy of including only smooth manifolds. The $\mathbb{Z}_N$ generators act on the $AdS_3$ bulk with fixed points, leading to a codimension-2 singularity in the interior and a deficit angle
\begin{equation}\label{eq:conicalDefects}
  \Delta\phi_N = 2\pi(1-N^{-1}).
\end{equation}
The metric of the $\mathbb{Z}_N$ orbifolds of thermal $AdS_3$, $\mathcal{M}_{0,1}/\mathbb{Z}_N$, is given by (\ref{eq:euclideanMetric}), but the quotient by $Y$ as in (\ref{eq:GammaGeneratorY}) leads to the further identification $\phi \sim \phi + {2\pi\over N}$.
See figure \ref{fig:singularBulk} for a cartoon of the Euclidean geometry.

The conventional wisdom regarding such singularities in three-dimensional gravity is that they signal the presence of massive particles with Planckian mass that backreact on the geometry and source a conical defect. In the classical limit, the deficit angle $\Delta\phi$ sourced by a massive scalar particle is related to its mass by \cite{Deser:1983tn}
\begin{equation}
  \Delta\phi \approx 8\pi G_N m.
\end{equation}
The masses $m_N$ corresponding to the $\mathbb{Z}_N$ conical defects lead to the following conformal weights in the classical limit
\begin{equation}\label{eq:largeCOrbifoldWeights}
  h_N \approx {c\over 24}(1-N^{-2}).
\end{equation}
In \cite{Collier:2018exn} a proposal was made for a (generically non-geometric) notion of conical defect states at finite central charge associated to local CFT operators, based on the observation that the large-spin spectrum is additive in the Liouville momentum $\alpha$, which corresponds in the classical limit to the additivity of the conical deficit angles at large separation. The proposal was\footnote{The Liouville variables $\alpha(h),Q(c)$ are defined below and in appendix \ref{app:regs}.}
\begin{equation}
  \Delta\phi = {4\pi\alpha\over Q} = 2\pi\(1-\sqrt{1-8G_Nm}\).
\end{equation}
This provides a plausible definition of the conical defect states in the quantum regime. The finite $c$ value of the reduced twist corresponding to the $\mathbb{Z}_N$ deficit angle (\ref{eq:conicalDefects}) is then given precisely by (\ref{eq:orbifoldTwist}), one of the values of the twist whose addition to the gravitational path integral may cure the large-spin non-unitarity of the pure gravity spectrum.

Here, we will consider such singular geometries as equally valid contributions to the gravitational path integral. In particular, we will consider the addition of states with reduced twist (\ref{eq:orbifoldTwist}) along with their $PSL(2,\mathbb{Z})$ images. This corresponds to including the singular orbifold saddles, with all possible linear combinations of the boundary cycles taken to be contractible, in the gravitational path integral with torus boundary.

\begin{figure}[h!]
  \centering
  \subfloat{\includegraphics[width=7cm]{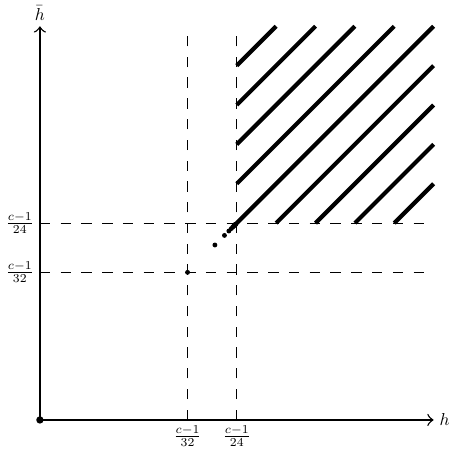}
  }
  \subfloat{\includegraphics[width=7cm]{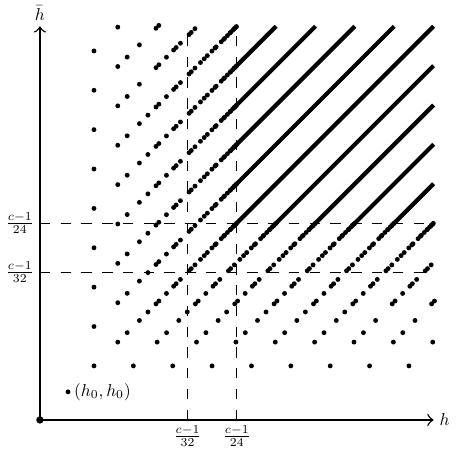}
  }
  \caption{\textbf{Left:} The spectrum of Virasoro primary operators of the pure gravity partition function endowed with the conical defect states. There are no discrete Regge trajectories with an asymptotic twist below the black hole threshold, and the spectrum of twists is continuous above the black hole threshold. \textbf{Right:} A rough sketch of the spectrum of Virasoro primary operators of a generic irrational CFT with a light scalar operator with weight $h_0$ below ${c-1\over 32}$. In this case there are a finite, discrete set of infinite towers of multi-twist operators with asymptotic twist below the black hole threshold. This sketch is meant to be schematic, and merely capture the existence of the discrete multi-twist Regge trajectories below the black hole threshold --- in a genuine CFT, the Regge trajectories are not exactly flat (i.e. the multi-twist operators have anomalous dimensions at finite spin) and the spectrum of twists above the black hole threshold is only continuous at large spin. Furthermore, the presence of the multi-twist Regge trajectories will serve to modify the black hole threshold in a spin-dependent way; see \cite{Maxfield:2019hdt} for more details.}
  \label{fig:contrast}
\end{figure}

Although such a modification to the gravitational path integral may violate the aesthetic ideals of ``pure'' quantum gravity, we would like to point out that the addition of these states is qualitatively different at the level of the spectrum of the dual CFT than the addition of a generic light operator, in the following sense. In \cite{Collier:2018exn} and \cite{Kusuki:2018wpa}, it was shown that an irrational CFT with $c>1$ and nonzero twist gap has the following property: if the spectrum includes two light Virasoro primary operators with twists $h_1$ and $h_2$, then the spectrum also contains infinitely many primary operators arranged into discrete Regge trajectories, with twists that accumulate to the following ``Virasoro double-twist'' values at large spin
\begin{equation}
  h_m = h_1+h_2+m+\delta h_m,~\text{for each integer $m\ge 0$ such that } h_m < {c-1\over 24},
\end{equation}
where
\begin{equation}\label{eq:deltahm}
  \delta h_m = -2(\alpha(h_1)+mb(c))(\alpha(h_2)+mb(c))+m(m+1)b^2(c) <0
\end{equation}
is the anomalous twist resulting from the exact summation of the exchange of all multi-stress tensor composites. Here, $\alpha(h)$ is a function of the weight defined in terms of the Liouville variables\footnote{See (\ref{eq:liouvilleMomenta}) and (\ref{eq:backgroundCharge}) for explicit definitions of $P,Q,b$ in terms of the usual CFT$_2$ variables.} as 
\begin{equation}
  \alpha(h) = {Q\over 2}+iP(h)\in\left(0,{Q\over 2}\right)\text{ for }0<h<{c-1\over 24}.
\end{equation}
In the bulk, these infinite towers of states have the interpretation of multi-particle bound states in $AdS_3$, with the large-spin anomalous dimension (\ref{eq:deltahm}) corresponding to the gravitational binding energy due to the totality of (multi-)graviton exchanges, which is nontrivial even at infinite separation in $AdS_3$. \emph{Precisely} above the value $h = {c-1\over 32}$ (corresponding to the $\mathbb{Z}_2$ orbifold seed state), there can be no such discrete towers of composite operators built out of the seed operator; this is because $2\alpha({c-1\over 32}) = {Q\over 2}$. Correspondingly, there are no infinite towers of multi-particle bound states built out of the particle that sources the conical defect. See figure \ref{fig:contrast} for a qualitative comparison between the pure gravity spectrum endowed with conical defect states and the spectrum of an irrational CFT with a generic light operator. We thus view our proposed modification as a minimal violation of the conceptual ideals of pure quantum gravity, in the sense that the new states have purely geometrical interpretations, and do not lead to infinite towers of multi-particle bound states in $AdS_3$.

We note in passing that the need to include subleading saddles corresponding to orbifold geometries in the near-horizon $AdS_2$ functional integral in order to reproduce non-perturbative corrections to the entropy of four-dimensional supersymmetric black holes from localization has been previously observed in \cite{Dabholkar:2014ema,Murthy:2009dq,Banerjee:2008ky}.\footnote{We are grateful to Xi Yin for bringing this to our attention.} However, these orbifold geometries are typically non-singular from the point of view of the full near-horizon geometry.

\section{The one-loop determinant on orbifold backgrounds}\label{sec:oneLoop}

We will now consider the perturbative corrections to the orbifold geometries described above. In this section we will study the one-loop partition function for gravitons on backgrounds that are $\mathbb{Z}_N$ orbifolds of thermal $AdS_3$, namely $\mathbb{H}_3/(\mathbb{Z}\times \mathbb{Z}_N)$, where $\mathbb{H}_3$ is hyperbolic three-space. For the purposes of this computation of the gravitational one-loop determinant of the conical defect geometries it is essential that $N$ is an integer. We will closely follow the approach of \cite{Giombi:2008vd}, where the one-loop gravity partition function on Euclidean $AdS_3$ was computed using the heat kernel method and the method of images. There, it was shown that the one-loop determinant on Euclidean $AdS_3$ yielded an answer that was consistent with the Virasoro vacuum character
\begin{equation}\label{eq:VirasoroVacuum}
  Z_{\text{1-loop}}(\tau,\bar\tau) = (q\bar q)^{-{c\over 24}}\prod_{m=2}^\infty{1\over |1-q^m|^2},\quad c = {3L\over 2G_N}+\mathcal{O}(1),
\end{equation}
where $L$ is the $AdS_3$ radius.

Since our backgrounds of interest are orbifolds of thermal $AdS_3$, we will simply apply the method of images to the results of \cite{Giombi:2008vd}. We expect to recover the non-degenerate Virasoro character
\begin{equation}\label{eq:expectation}
  Z^{(N)}_{1\text{-loop}}(\tau,\bar\tau) \stackrel{?}{=}{(q \bar q)^{T-{1\over 24}}}\prod_{m=1}^\infty{1\over|1-q^m|^2},
\end{equation}
where the reduced twist $T$ ought to agree with the values (\ref{eq:orbifoldTwist}) and (\ref{eq:orbifoldTildeTwist}) required to render the pure gravity spectrum positive, at least to leading order in $c$.
We emphasize that equation (\ref{eq:expectation}) is one-loop exact, as it is completely fixed by the central charge and the structure of the Virasoro algebra.  Thus, although we  will not explicitly compute the two-loop and higher corrections, we expect that their only effect will be to renormalize the effective value of the central charge.

\subsection{Review of the gravitational one-loop determinant on thermal $AdS_3$}
We will start by briefly reviewing the computation of \cite{Giombi:2008vd} for thermal $AdS_3$. The heat kernel $K(t;x,x')$ is a function of two spacetime positions $x,x'$ and an auxiliary time variable $t$. Suppose one is interested in computing the one-loop contribution to a partition function $Z$
\begin{equation}
  Z = \int D\phi\, \exp\({-h^{-2}\int d^d x\sqrt{g}\,\phi\Delta\phi}\),
\end{equation}
where $\Delta$ is a Laplacian and $h$ is a perturbative coupling. The heat kernel is typically defined in terms of eigenfunctions of the Laplacian, $\Delta_x\phi_n(x) = \lambda_n\phi_n(x)$ as
\begin{equation}
  K(t;x,x') = \sum_n e^{-\lambda_n t}\phi_n(x)\phi_n(x'),
\end{equation}
where the eigenfunctions $\phi_n$ have been normalized in a convenient way, $\sum_n\phi_n(x)\phi_n(x') = \delta^d(x-x')$ and $\int d^d x\sqrt{g}\,\phi_n(x)\phi_m(x) = \delta_{nm}$. The one-loop contribution to the partition function is computed by the logarithm of the determinant of the Laplacian\footnote{This simplified discussion does not directly apply to the case of interest because the relevant geometries are non-compact, leading to a divergent contribution to $S^{(1)}$ proportional to the volume of spacetime. Furthermore in the non-compact case the Laplacian typically does not have a discrete spectrum of eigenvalues.}
\begin{equation}
  S^{(1)} = -\half\sum_n\log\lambda_n.
\end{equation} 
This quantity can be conveniently computed by the trace of the heat kernel $K$
\begin{equation}
  S^{(1)} = \half\int_0^\infty{dt\over t}\int d^3 x\sqrt{g}\,K(t;x,x).
\end{equation}

Luckily for us, the heat kernels for gravitons on $\mathbb{H}_3$ and $\mathbb{H}_3/\ZZ$ have been computed by \cite{Giombi:2008vd}, and we may simply proceed to use the method of images in order to compute the gravitational one-loop determinant on the orbifold backgrounds. The one-loop determinant on $\mathbb{H}_3/\ZZ$ takes the form \cite{Giombi:2008vd}
\begin{equation}\begin{aligned}\label{eq:thermalAdsOneLoop}
  S^{(1)} =& \vol(\mathbb{H}_3/\ZZ)\int_0^\infty{dt\over t}{1\over 4\pi t^{3/2}}\(e^{-t}(1+8t)-e^{-4t}(1+2t)\)\\
  &+\sum_{n\ne 0}\int_0^\infty{dt\over t}\int_{\mathbb{H}_3/\ZZ}d^3 x\sqrt{g}\,K^{\mathbb{H}_3}(t,r(x,\gamma^n x)).
\end{aligned}\end{equation}
Many comments are in order. The first is that the first term is divergent and will require regularization; there is an IR divergence due to the infinite volume of $\mathbb{H}_3/\ZZ$ and a UV divergence due to the $t\to 0$ behaviour of the integrand, both of which can be dealt with by a local counterterm. Here, $\ZZ$ is generated by $\gamma$, which acts on $\mathbb{H}_3$, parameterized by the metric
\begin{equation}
  ds^2 = {dy^2+dzd\bar z\over y^2},
\end{equation}
as
\begin{equation}
  \gamma(y,z,\bar z)  = (|q|^{-1}y,q^{-1}z,\bar q^{-1}\bar z),
\end{equation}
where $q = e^{2\pi i\tau}$ is the modulus of the torus boundary. The geodesic distance $r$ that enters the heat kernel is given by
\begin{equation}
  r(x,x') = \text{arccosh}\left(1+{(y-y')^2+|z-z'|^2\over 2y y'}\right).
\end{equation} 
It is often convenient to make use of polar coordinates
\begin{equation}\begin{aligned}
  y &= \rho\sin\theta\\
  z &= \rho\cos\theta e^{i\phi}.
\end{aligned}\end{equation}
These coordinates range from $1\le \rho< e^{2\pi\tau_2}$, $0\le\phi<2\pi$ and $0\le \theta<{\pi\over 2}$ on $\mathbb{H}_3/\ZZ$. In terms of these coordinates, the geodesic distance in the second term in (\ref{eq:thermalAdsOneLoop}) is given by
\begin{equation}
  r(x,\gamma^n x') = \arccosh\(1+2\sinh^2(n\pi\tau_2)+2\cot^2\theta|\sin(n\pi\tau)|^2\),
\end{equation}
and the corresponding volume element is given by
\begin{equation}
  d^3x\sqrt{g} = d\phi {d\rho\over \rho}{dr\,\sinh r\over 4|\sin\pi n \tau|^2}.
\end{equation}

The end result of the computation in \cite{Giombi:2008vd}, the details of which we will not attempt to reproduce here, is the following for the gravitational one-loop determinant on thermal $AdS_3$
\begin{equation}\begin{aligned}\label{eq:thermalAdsDeterminant}
  S^{(1)} =& \vol(\mathbb{H}_3/\ZZ)\int_0^\infty{dt\over t}{1\over 4\pi t^{3/2}}\(e^{-t}(1+8t)-e^{-4t}(1+2t)\)\\
  &+\int_0^\infty{dt\over t}\sum_{n=1}^\infty{2\pi^2\tau_2\over|\sin \pi n\tau |^2}{e^{-{(2\pi n\tau_2)^2\over 4t}}\over 4\pi^{3/2}\sqrt{t}}\(e^{-t}\cos 4\pi n\tau_1-e^{-4t}\cos 2\pi n\tau_1\)\\
  =& -{13\over 6\pi}\vol(\mathbb{H}_3/\ZZ) - \sum_{j=2}^\infty\log |1-q^j|^2.
\end{aligned}\end{equation}
The first term is computed by analytically continuing the integral over $t$ by using the integral representation of the gamma function. After holographically renormalizing the infinite volume of $\mathbb{H}_3/\ZZ$ and combining with the classical piece, the one-loop gravity partition function on $\mathbb{H}_3/\ZZ$ then takes the form of the Virasoro vacuum character (\ref{eq:VirasoroVacuum}) with
\begin{equation}\label{eq:oneLoopCentralCharge}
  c = {3L\over 2G_N}+13.
\end{equation}
This one-loop shift of the central charge has recently been emphasized in \cite{Cotler:2018zff}. Although it can be removed by a local counterterm, we will keep it for the purposes of comparison to the orbifold computation, where we will no longer have the freedom to renormalize the cosmological constant.

\subsection{The gravitational one-loop determinant on the conical defect backgrounds}
The heat kernel on $\mathbb{H}_3/(\mathbb{Z}\times \mathbb{Z}_N)$ can be obtained from that on $\mathbb{H}_3$ by the method of images
\begin{equation}
  K^{\mathbb{H}_3/(\mathbb{Z}\times \mathbb{Z}_N)}(t,x,x') = \sum_{m=1}^N\sum_{n\in\mathbb{Z}}K^{\mathbb{H}_3}(t,r(x,\gamma^{n,m}_Nx')),
\end{equation}
where
\begin{equation}
	\gamma^{n,m}_{N}(y,z,\bar z) \to (|q|^{-n/N}y, q^{-n/N}e^{-2\pi i m/N }z,\bar q^{-n/N}e^{2\pi i m/N }\bar z),
\end{equation}
where  $n\in\ZZ$ and $m\in\{1,2,\ldots,N\}$. Note that throughout this computation we have rescaled $\tau$ by $N^{-1}$ so that the complex structure of the boundary torus is the same as in the $N=1$ case. The geodesic distance that enters the heat kernel via the method of images then takes the form
\begin{equation}
  r(x,\gamma^{n,m}_N x) =\arccosh\(1+2\sinh^2\( {\pi n\tau_2\over N}\)+2\cot^2\theta\left|\sin\({\pi\over N}(n\tau+m)\)\right|^2\).
\end{equation}

We are now well-positioned to evaluate the gravity one-loop determinant on $\mathbb{H}_3/(\ZZ\times \ZZ_N)$, which is given by
\begin{equation}\begin{aligned}
  S^{(1)} =& \vol(\mathbb{H}_3/(\ZZ\times\ZZ_N))\int_0^\infty{dt\over t}{1\over 4\pi t^{3/2}}\(e^{-t}(1+8t)-e^{-4t}(1+2t)\)\\
  &+\sum_{m=1}^{N-1}\int_0^\infty{dt\over t}\int_{\mathbb{H}_3/(\ZZ\times \ZZ_N)}d^3 x\sqrt{g}\,K^{\mathbb{H}_3}(t,r(x,\gamma_N^{0,m} x))\\
  &+\sum_{m=1}^N\sum_{n\ne 0}\int_0^\infty{dt\over t}\int_{\mathbb{H}_3/(\ZZ\times\ZZ_N)}d^3 x\sqrt{g}K^{\mathbb{H}_3}(t,r(x,\gamma_N^{n,m}x))\\
  \equiv&\, \delta_1+\delta_2+\delta_3,
\end{aligned}\end{equation}
where
\begin{equation}
\begin{aligned}
  \delta_1 &= -{13\over 6\pi}\vol(\mathbb{H}_3/(\ZZ\times\ZZ_N))\\
	\delta_2 &= \half\sum_{m=1}^{N-1}\int_0^\infty {dt\over t}{2\pi^2\tau_2\over N^2\sin(\pi m/N)^24\pi^{3/2}\sqrt{t}}\left[e^{-t}\cos(4\pi m/N)-e^{-4t}\cos(2\pi m/N)\right]\\
	\delta_3 &= \half\sum_{m=1}^N\sum_{n=1}^\infty\int_0^\infty{dt\over t}{2\pi^2\tau_2 e^{-{(2\pi n\tau_2)^2\over 4N^2t}}\over 4\pi^{3/2}N^2\sqrt{t}}\left[{e^{-t}\cos\({4\pi\over N}(n\tau_1+m)\)-e^{-4t}\cos\({2\pi\over N}(n\tau_1+m)\)\over|\sin\({\pi\over N}(n\tau+m)\)|^2}+(n\to-n)\right].
\end{aligned}
\end{equation}
Both $\delta_1$ and $\delta_2$ are formally divergent. The divergence in $\delta_1$ is due to the non-compactness of spacetime, while the divergence in $\delta_2$ from the $t\to0$ part of the integrand is due to the fixed points of the $\ZZ_N$ action. As we will see, both of these divergences can be regularized, although in the latter case we do not have an a priori physical justification for the choice of regularization scheme. 

We will start by considering the final term. Borrowing some manipulations from \cite{Giombi:2008vd}, it can be written as the following
\begin{equation}
\begin{aligned}\label{eq:delta3}
	\delta_3 &= \sum_{m=1}^N\sum_{n=1}^\infty{1\over 2nN}\left[{e^{4\pi i m/N}q^{2n\over N}\over 1-e^{2\pi im/N}q^{n\over N}}+{e^{-4\pi im/N}\bar q^{2n\over N}\over 1-e^{-2\pi im/N}\bar q^{n\over N}}+(m\to-m)\right]\\
	&=\sum_{m=1}^N\sum_{n=1}^\infty\sum_{\ell=0}^\infty{1\over nN} \cos(2\pi m(\ell+2)/N)\left(q^{n(\ell+2)\over N}+\bar q^{n(\ell+2)\over N}\right)\\
	&= \sum_{n=1}^\infty\sum_{\ell=0}^\infty{\delta^{(N)}_{\ell+2,0}\over n}\left(q^{n(\ell+2)\over N}+\bar q^{n(\ell+2)\over N}\right)\\
	&= -\sum_{j =1}^\infty\log|1-q^{j}|^2,
\end{aligned}
\end{equation}
where $\delta_{m,n}^{(N)} = 1$ if $m \equiv n ~(\text{mod}~N)$, and vanishes otherwise. This is precisely of the form of the non-degenerate Virasoro character that encodes contributions from descendants on a torus with modular parameter $q$. Notice that the effect of the sum over $\ZZ_N$ images is to produce a sum that starts at $j=1$ rather than $j=2$ as in (\ref{eq:thermalAdsDeterminant}).

Let's now move on to the second term, for which the $t$ integral will require regularization, due to the fixed points in the $\ZZ_N$ action. As in the $\mathbb{H}_3/\ZZ$ example, we will analytically continue using the integral representation of the gamma function. We have
\begin{equation}
\begin{aligned}\label{eq:delta2}
	\delta_2  &= -\sum_{m=1}^{N-1}{\half\pi\tau_2\over N^2\sin(\pi m/N)^2}[\cos(4\pi m/N)-2\cos(2\pi m/N)]\\
	&=+4\pi \tau_2 {N^2-13\over 24 N^2}.
\end{aligned}
\end{equation}
This almost takes the form of the one-loop correction to the dimension for scalar operators with reduced twist $T=T_N$, except the shift in the numerator is by $-13$ instead of the expected $-1$. However, this contribution from the $\mathbb{Z}_N$ images of the singular locus is potentially ambiguous due to the need to regularize, and may be subject to further ambiguities in the choice of boundary conditions for the graviton at the singular locus.\footnote{We thank Hirosi Ooguri, Yifan Wang, and Xi Yin for discussions on this point.} Although we do not have an a priori physical justification for our choice of regularization scheme, we simply choose to regularize in the same way as in thermal AdS$_3$, where the analytic continuation of the integral representation of the gamma function yields the one-loop correction to the central charge given in (\ref{eq:oneLoopCentralCharge}).

Combining the one-loop results (\ref{eq:delta3}) and (\ref{eq:delta2}) with the classical action and renormalizing the volume of the orbifold spacetime as in the thermal $AdS_3$ computation, we arrive at the following expression for the gravitational one-loop partition function on the orbifold backgrounds\footnote{The result of a version of the computation in this section was also presented in \cite{Oblak:2016eij}, however there the one-loop correction to the conformal weight was not included.}
\begin{equation}
\begin{aligned}
  Z^{(N)}_{\text{1-loop}} &= e^{-({3L\over 2G_N}+13){1\over 6\pi}\vol(\mathbb{H}_3/(\mathbb{Z}\times\mathbb{Z}_N))} ( q{\bar{ q}})^{-{N^2-13\over 24 N^2}}\prod_{m=1}^\infty{1\over |1- q^m|^2}\\
  &= ( q{\bar{ q}})^{-{{3L\over 2 G_N}+13\over 24 N^2}-{N^2-13\over 24N^2}}\prod_{m=1}^\infty{1\over |1- q^m|^2}\\
  &= {(q{\bar{ q}})^{-{1\over 24 N^2}{3L\over 2G_N}}\over |\eta( \tau)|^2},
\end{aligned}
\end{equation}
where, following \cite{Cotler:2018zff}, we have renormalized the volume of spacetime as
\begin{equation}
  \text{vol}(\mathbb{H}_3/(\mathbb{Z}\times\mathbb{Z}_N)) = -{\pi^2\tau_2\over N^2}.
\end{equation}
Given that the contribution from the thermal $AdS_3$ saddle is a Virasoro vacuum character with central charge $c = {3L\over 2G_N}+13$, this suggests that the reduced twist of the primary corresponding to the Virasoro character we obtain from computing the gravitational path integral over the singular orbifold geometries is given by
\begin{equation}
 	T =  - {c-13\over 24N^2} = -{c-1\over 24 N^2} + {1\over 2N^2}.
	\label{eq:TOneLoop}
\end{equation}
We have arrived at the result that the twist of the corresponding non-degenerate Virasoro character appears to be halfway between $T_N$ and $\widetilde T_N$. We do not have a physical explanation for this result, although it is perhaps suggestive that the resulting twist is the arithmetic mean of the two values of the twist whose inclusion in the partition function is suggested by the modular bootstrap argument.

\section{Discussion} 
\label{sec:discuss}

In this paper we have reviewed the torus partition function of pure three-dimensional gravity. We have described a natural modification to \cite{Maloney:2007ud, Keller:2014xba} that is positive everywhere that maximizes the dimensions of extra operators added while minimizing their degeneracies. These operators added have the correct leading behavior in $c$ to be interpreted as singular orbifolds of $AdS_3$ by $\mathbb Z_N$. We pause here to make a few comments and list interesting questions.

The first comment we make is that the spectrum of the resulting theory is continuous. Moreover, the one-loop correction in a $1/c$ expansion of the orbifold geometries does not precisely match those of the operators that we add. A conservative viewpoint would be that these facts are evidence that a pure theory of gravity in $AdS_3$ does not exist, even when these singular orbifolds are included in a sum over geometries. Indeed this is similar to the philosophy advocated in \cite{Maloney:2007ud}. An alternative possibility is that the spectrum we write down is not dual to a single CFT, but rather an ensemble average of CFTs in a similar spirit as \cite{Saad:2019lba, Stanford:2019vob}. This would allow for a continuous density of states at each spin that is in an appropriate sense the ``averaged value" of the partition functions of the ensemble of CFTs. We also note that the regularized one-loop correction to the orbifold saddles is a non-degenerate Virasoro character with a twist (\ref{eq:TOneLoop}) that is precisely the average value of the two possible twists whose addition is suggested by the modular bootstrap argument: $T_N, \widetilde T_N$ in (\ref{eq:orbifoldTwist}) and (\ref{eq:orbifoldTildeTwist}), although the precise order-one value of the twist may be subject to ambiguities due to the need to regularize. 

The second comment we make is about how many orbifold geometries to include in the sum over geometries. As we saw in section \ref{sec:minimalist}, unitarity only requires a finite number of orbifold geometries to include. One possibility, then, is that only a finite number of orbifolds are included in the path integral sum, since they are all that are needed to cancel the negativities in the spinning sector of the partition function. On the other hand, it may seem unnatural to only include some but not all of the orbifold geometries. If we include all orbifold geometries and interpret them as states with zero spin, there would be an accumulation point in dimension at finite $c$. This may be an indication that some regularization procedure of these states can be used to cancel the delta function negativity at $h=\bar{h}=\frac{c-1}{24}$ from the $PSL(2,\mathbb Z)$ sum (which currently is removed via the rather ad hoc procedure of adding six compact free boson partition functions). Another possibility is that there is a $c$-dependent cutoff for the number of orbifolds one needs to include in the sum, with the number of orbifolds included going to infinity in the limit of large $c$. A third possibility is we give each orbifold an $\mathcal{O}(1)$ spin. For instance for each state $T_N$ and $\widetilde T_N$, we include two copies: one at spin $N-1$ and one at spin $-(N-1)$, and cut off $N$ as a function of $c$. This would both remove the negativity at $h=\bar{h}=\frac{c-1}{24}$, and avoid the accumulation point at finite energy.

Let us comment more on the sum over orbifolds. Remarkably, if one includes all the orbifold geometries (as scalars), even though there is an accumulation point at finite dimension, the partition function remains finite. The reason is the following. Consider the $PSL(2,\mathbb Z)$ sum of a nondegenerate Virasoro character at reduced twist $T<0$ and spin $J=0$:
\be
\sum_{\gamma \in\mathbb Z \backslash PSL(2,\mathbb Z)}\frac{(q_\gamma\bar{q}_\gamma)^T}{\eta(q_\gamma)\eta(\bar q_\gamma)}.
\label{eq:psl2zsumg}
\ee
In (\ref{eq:psl2zsumg}), the sum over $PSL(2,\mathbb Z)$ should be interpreted with the regularization of \cite{Maloney:2007ud} which we reviewed in section \ref{sec:ReviewMWK} and appendix \ref{app:regs}. Performing this sum (i.e. taking the Laplace transform of (\ref{eq:generalPoincare}) and (\ref{eq:regularizedPoincareScalar})) gives:\footnote{See also sections 3.1 and 3.2 of \cite{Keller:2014xba}.}
\begin{align}
	Z_{T}(\tau,\bar\tau) =& \frac1{\eta(q)\eta(\bar q)} \Bigg((q\bar q)^T - 1 +  \sum_{m=1}^{\infty} \frac{\zeta(2m)\Gamma(m)(-16\pi T)^m}{\zeta(2m+1)(2m)!\tau_2^m}  \nonumber\\&+\sum_{j=1}^{\infty} 2\(e^{2\pi i j\tau_1} + e^{-2\pi i j \tau_1}\)\sum_{m=1}^\infty  \frac{(-16\pi^2 T)^m \sigma_{2m}(j)}{(2m)! j^m \zeta(2m+1)} K_m(2\pi j\tau_2)\Bigg),
	\label{eq:MWKnoNull}
\end{align}
where $\sigma_n(j) = \sum_{d|j}d^n$ is the divisor function and $K_m$ is the modified Bessel function of the second kind.

If we do a sum over orbifolds for each $\mathbb Z_N$, then we will essentially be summing (\ref{eq:psl2zsumg}) with $T$ scaling as 
\be
T^{(N)} = \frac1{N^2} \(-\frac{c}{24} + \mathcal{O}(1)\)
\label{eq:en}
\ee
where the $\mathcal{O}(1)$ correction in (\ref{eq:en}) is a $c$-independent piece that depends on if we take $T_N$, $\widetilde T_N$, etc. that will be irrelevant for the following discussion. It turns out that (\ref{eq:MWKnoNull}) at small $T$ and finite $\tau,\bar\tau$, the $T$-dependence in the sum (\ref{eq:psl2zsumg}) scales linearly with $T$:
\be
Z_{T}(\tau, \bar\tau) \sim T,~~~~T~\text{small}.
\label{eq:scalingZ}
\ee
The intuition for (\ref{eq:scalingZ}) is that the first and second term in (\ref{eq:MWKnoNull}) cancel at small $T$. Thus the negativity from the delta function is key to arrive at (\ref{eq:scalingZ}). At large $N$, the partition function sum scales as 
\be
\sum_{N=1}^\infty Z_{T^{(N)}}(\tau, \bar\tau) \sim \sum_{N=1}^\infty \frac1{N^2}\(-\frac{c}{24} + \mathcal{O}(1)\)
\label{eq:asdfgh}
\ee
which is finite (in (\ref{eq:scalingZ}), (\ref{eq:asdfgh}) we are suppressing the $\tau,\bar\tau$ dependence).  This means that the partition function after summing over all orbifold geometries is finite, despite the accumulation point in dimension!\footnote{Interestingly if one repeats the same calculation but instead uses the Rademacher expansion rather than the Poincar\'e series, corresponding to a slightly different ``seed'' spectrum, as in \cite{Alday:2019vdr}, the partition function at small $T$ would scale as $\sqrt{T}$, not $T$, which would lead to a divergence in the sum over $N$.} Roughly, the divergence due to the accumulation point at extremality is cancelled by the infinite number of ``$-1$'' subtractions (the second term in (\ref{eq:MWKnoNull}), corresponding to the negative delta function precisely at extremality arising from the zeta function regularization as discussed in appendix \ref{app:regs}).

In fact, the partition function resulting from the Poincar\'e series of all of the $\ZZ_N$ conical defect states with reduced twist $T_N$ can be written in the following compact way
\begin{align}
\sum_{N=1}^\infty Z_{T_N}(\tau,\bar\tau) =& \frac1{\eta(q)\eta(\bar q)} \Bigg(\sum_{m=1}^\infty{(4\pi\xi\tau_2)^m\over m!}\zeta(2m) +  \sum_{m=1}^{\infty} \frac{\Gamma(m)(16\pi \xi)^m\zeta(2m)^2}{\zeta(2m+1)(2m)!\tau_2^m}  \nonumber\\&+\sum_{j=1}^{\infty} 2\(e^{2\pi i j\tau_1} + e^{-2\pi i j \tau_1}\)\sum_{m=1}^\infty  \frac{(16\pi^2 \xi)^m \sigma_{2m}(j)\zeta(2m)}{(2m)! j^m \zeta(2m+1)} K_m(2\pi j\tau_2)\Bigg),
\label{eq:sumOverOrbifolds}
\end{align}
where $\xi = {c-1\over 24}$. Similarly the sum over $\widetilde T_N$ is given by the same expression (\ref{eq:sumOverOrbifolds}) but with $\xi$ replaced with $\xi-1 = \frac{c-25}{24}$. Finally we subtract out the $PSL(2,\mathbb{Z})$ sum of a seed spectrum with $T = -\frac{c-1}{24}, J=\pm 1$, corresponding to the null descendants of the identity (see the results in sections 3.3 and 3.4 of \cite{Keller:2014xba}, which we rewrite for convenience in appendix \ref{app:fullorbifoldspectrum}).

It would be extremely interesting if there were a physical interpretation for the finiteness of this partition function, perhaps as an ensemble average of CFT partition functions since it does not appear to admit a standard decomposition into Virasoro characters. Moreover the density of states when integrated against a test function with support at $h=\bar{h}=\frac{c-1}{24}$ can become negative, since we have not added the free boson partition functions in order to preserve finiteness.\footnote{We thank H. Maxfield for emphasizing this point.} We leave this question for future work. 

Going forward, one ought to study not just the spectrum but also the dynamics of pure gravity in $AdS_3$. The extreme sparseness of the light spectrum of the quantum theory of pure gravity means that the universal asymptotic formulas for OPE data recently derived in \cite{Collier:2019weq} (corrections to which are controlled by the low-lying operators in the spectrum) ought to have a maximally-extended regime of validity in the large central charge limit, analogous to the case of the Cardy formula studied in \cite{Hartman:2014oaa}. We also note that the universal structure constant $C_0(h_1,h_2,h_3)$ can be straightforwardly analytically continued for conformal weights precisely as low as $h_i = {c-1\over 32}$. Perhaps the most straightforward route to access the structure constants would be to study the pure gravity genus-two partition function. We leave this interesting line of inquiry for future study.

\section*{Acknowledgements}
We would like to thank L. F. Alday, J. Cotler, K. Jensen, H. Maxfield, H. Ooguri, S.-H. Shao, H. Verlinde, Y. Wang, and X. Yin for interesting discussions.  
We thank H. Maxfield, H. Ooguri, S.-H. Shao, and Y. Wang for detailed comments on a draft.
We are grateful to the organizers of the 2019 meeting of the Simons Collaboration on the Nonperturbative Bootstrap at Perimeter Institute, during which this work was instigated. Research at Perimeter Institute is supported in part by the Government of Canada through the Department of Innovation, Science and Economic Development Canada and by the Province of Ontario through the Ministry of Economic Development, Job Creation and Trade. 
The work of N.B. is supported in part by the Simons Foundation Grant No. 488653. 
The work of A.M. is supported in part by the Simons Foundation Grant No. 385602 and the Natural Sciences and Engineering Research Council of Canada (NSERC), funding reference number SAPIN/00032-2015.
This research was supported in part by the National Science Foundation under Grant No. NSF PHY-1748958.

\appendix

\section{MWK regularization} 
\label{app:regs}

In this appendix we will show that the regularized sum over modular kernels (\ref{eq:sumgamma}) precisely reproduces the MWK spectrum. We start with (\ref{eq:sumgamma}) and do some manipulations:
\begin{equation}
\begin{aligned}\label{eq:MWKDensity}
  \rho^{\rm MWK}(h,\bar h) =& \sum_{\gamma\in \mathbb Z \backslash PSL(2,\ZZ)}\left[\kernel^{(\gamma)}_{h0}\overline{\kernel}^{(\gamma)}_{\bar h 0}-\kernel^{(\gamma)}_{h0}\overline{\kernel}^{(\gamma)}_{\bar h 1} -\kernel^{(\gamma)}_{h1}\overline{\kernel}^{(\gamma)}_{\bar h 0} + \kernel^{(\gamma)}_{h1}\overline{\kernel}^{(\gamma)}_{\bar h 1}\right]\\
  =& \sum_{s=1}^\infty\sum_{d'\in(\ZZ/s\ZZ)^*}\sum_{n=-\infty}^{\infty}{2\over sP\bar P}e^{2\pi i(d'+ns) j\over s}\bigg[\cosh{2\pi QP\over s}\cosh{2\pi Q\bar P\over s} - e^{-{2\pi i (d'^{-1})_s\over s}}\cosh{2\pi QP\over s}\cosh{2\pi\widetilde Q\bar P\over s}\\
  &-e^{{2\pi i(d'^{-1})_s \over s}}\cosh{2\pi \widetilde Q P\over s}\cosh{2\pi Q\bar P\over s}+\cosh{2\pi \widetilde Q P\over s}\cosh{2\pi \widetilde Q\bar P\over s}\bigg]\\
  =& \sum_{\ell=-\infty}^\infty\sum_{s=1}^\infty{2\over sP\bar P}\delta(j-\ell)\bigg[S(j,0;s)\cosh{2\pi QP\over s}\cosh{2\pi Q\bar P\over s}-S(j,-1;s)\cosh{2\pi QP\over s}\cosh{2\pi\widetilde Q\bar P\over s}\\
  &-S(j,1;s)\cosh{2\pi \widetilde Q P\over s}\cosh{2\pi Q\bar P\over s}+S(j,0;s)\cosh{2\pi \widetilde QP\over s}\cosh{2\pi \widetilde Q\bar P\over s}\bigg],
\end{aligned}
\end{equation}
where the Liouville momenta $P,\bar P$ are defined in terms of the weights $h,\bar h$ as
\begin{equation}\label{eq:liouvilleMomenta}
  P = \sqrt{h-\frac{c-1}{24}}, ~~~ \bar P = \sqrt{\bar h-\frac{c-1}{24}},
\end{equation}  
and the background charge is defined in terms of the central charge as
\begin{equation}\label{eq:backgroundCharge}
  Q(c) = b(c)+b^{-1}(c) =  \sqrt{\frac{c-1}{6}}, ~~~~\widetilde Q(c) = b(c) - b^{-1}(c) =  \sqrt{\frac{c-25}{6}}.
\end{equation}
Finally, $S(j,J;s)$ is the Kloosterman sum, as in (\ref{eq:kloos}), and $j=h-\bar h$ is the spin. The sum is over $PSL(2,\ZZ)/\ZZ$ because the seed, the Virasoro vacuum character (\ref{eq:vacuumCharacter}), is independent of integer shifts of $\tau$. In the second line, following \cite{Maloney:2007ud}, we have decomposed the sum over $d$ by writing $d = d' + ns$, for $n\in\ZZ$ and $d'\in(\ZZ/s\ZZ)^*$, which denotes the subset of $\ZZ/s\ZZ$ with a multiplicative inverse. Decomposing into spin sectors, we reproduce (\ref{eq:MWKDensity2}), which we rewrite below for convenience:
\begin{equation}\begin{aligned}\label{eq:RewriteMWKDensity2}
 \rho^{\rm MWK}_{j}(t)= {2\over \sqrt{t(t+j)}}\sum_{s=1}^\infty{1\over s}&\bigg[S(j,0;s)\cosh\({{4\pi \over s}\sqrt{\frac{c-1}{24} (t+j)}}\)\cosh\({{4\pi \over s}\sqrt{\frac{c-1}{24} t}}\)  
  \\ &-S(j,-1;s)\cosh\({{4\pi\over s}\sqrt{\frac{c-1}{24} (t+j)}}\)\cosh\({{4\pi\over s}\sqrt{\frac{c-25}{24} t}}\)
  \\& -S(j,1;s)\cosh\({{4\pi \over s}\sqrt{\frac{c-25}{24}(t+j)}}\)\cosh\({{4\pi \over s}\sqrt{\frac{c-1}{24} t}}\) 
  \\&+S(j,0;s)\cosh\({{4\pi\over s}\sqrt{\frac{c-25}{24} (t+j)}}\)\cosh\({{4\pi\over s} \sqrt{\frac{c-25}{24} t}}\)\bigg].
\end{aligned}\end{equation}

To do the sum over $s$ we will need to regularize the modular kernel. Implicit in the definition of the modular kernel (\ref{eq:sl2zKernel}) is the fact that the Dedekind eta function transforms with weight one-half under $PSL(2,\mathbb{Z})$. For the purposes of analytic continuation, we now introduce a regularized crossing kernel corresponding to modular forms of generic weight $w$. In this case, the appropriate crossing relation is
\begin{equation}
  (-i(s\tau+d))^{-w}e^{2\pi i(\gamma\tau)(h-\frac{c-1}{24})} = \int_{\frac{c-1}{24}}^\infty dh'\,\mathbb{K}^{(w,\gamma)}_{h'h}e^{2\pi i \tau(h'-\frac{c-1}{24})},
\end{equation}
where the regularized modular crossing kernel is given by 
\begin{equation}\label{eq:regularizedKernel}
  \mathbb{K}^{(w,\gamma)}_{h'h} = \epsilon(w,\gamma)\({2\pi\over s}\)^w\(h'-\frac{c-1}{24}\)^{w-1} e^{{2\pi i\over s}(a(h-\frac{c-1}{24})+d(h'-\frac{c-1}{24}))}{{}_0F_1\(w;\frac{4\pi^2(\frac{c-1}{24}-h)(h'-\frac{c-1}{24})}{s^2}\)\over \Gamma(w)},
\end{equation}
with $w=\half$ being the case of physical interest. $\epsilon(w,\gamma)$ is again a phase that will be unimportant in what follows.

Consider, for example, the regularized contribution to the scalar ($j=0$) density of states of a scalar ($J=0$) seed
\begin{equation}\begin{aligned}\label{eq:regularizedPoincareScalar}
  \tilde\rho_{j=0}^{(T,J=0)}(t) &= \lim_{w\to\half}\sum_{s=1}^\infty\({2\pi\over s}\)^{2w}t^{2w-2}\Gamma(w)^{-2}S(0,0;s)+\sum_{s=1}^\infty{2\over s t}\left[\cosh^2\({4\pi\over s}\sqrt{-Tt}\)-1\right]S(0,0;s)\\
  &= \lim_{w\to\half}(2\pi)^{2w}{t}^{2w-2}{\zeta(2w-1)\over\zeta(2w)\Gamma(w)^2}+\sum_{s=1}^\infty{2\over s t}\left[\cosh^2\({4\pi\over s}\sqrt{-Tt}\)-1\right]S(0,0;s)\\
  &= -\delta(t)+\sum_{s=1}^\infty{2\over s t}\left[\cosh^2\({4\pi\over s}\sqrt{-Tt}\)-1\right]S(0,0;s),
\end{aligned}\end{equation}
where we obtained the delta function coefficient by integrating $t$ from $0$ to any finite positive number, and then taking the limit in $w$. This reproduces the negative delta function at $h=\bar h = \frac{c-1}{24}$, and the remaining sum converges in $s$. Similarly, the regularized contribution to the scalar ($j=0$) density of states from a seed state with reduced twist $T<0$ and spin $J>0$ is given by
\begin{equation}\begin{aligned}
  \tilde\rho_{j=0}^{(T,J)}(t) =& \lim_{w\to\half}\sum_{s=1}^\infty \({2\pi\over s}\)^{2w}{t}^{2w-2}\Gamma(w)^{-2}S(0,J;s)\\
  &+\sum_{s=1}^\infty{2\over s t}\left[\cosh\({4\pi\over s}\sqrt{-Tt}\)\cosh\({4\pi\over s}\sqrt{-(T+J)t}\)-1\right]S(0,J;s)\\
  =& 2\sigma_0(J)\delta(t)+\sum_{s=1}^\infty{2\over s t}\left[\cosh\({4\pi\over s}\sqrt{-Tt}\)\cosh\({4\pi\over s}\sqrt{-(T+J)t}\)-1\right]S(0,J;s).
\end{aligned}\end{equation}
Thus we see that the regularized MWK density of states
\begin{equation}
  \tilde\rho^{\rm MWK}_j(t) = \tilde\rho_j^{(T=-\frac{c-1}{24},J=0)}(t)-\tilde\rho_j^{(T=-\frac{c-1}{24},J=1)}(t)-\tilde\rho_j^{(T=-\frac{c-1}{24},J=-1)}(t)+\tilde\rho_j^{(T=-\frac{c-25}{24},J=0)}(t)
\end{equation}
contains a discrete negative degeneracy at extremality in the scalar sector
\begin{equation}
  \tilde\rho^{\rm MWK}_{j=0}(t) = -6\delta(t) +\text{(a continuous function of $t$)}.
\end{equation}

\section{A systematic analysis of the large-spin negativities}
\label{app:howmany}

All of the states we will need to add are at reduced twist $T_n = -\frac{c-1}{24n^2}$ and $\widetilde T_n =T_n + \frac{1}{n^2}$.  Let us now calculate how many states we will need to add for each $n$.

\subsection*{$\mathbf{n=2}$}

Since $2$ is prime the only term we need to worry about is the term at twist $0$, aka the vacuum. The term
\be
\frac{S(j, 0;2) - S(j, -1;2)}2 
\ee
is most negative for $j\equiv1~(\text{mod}~2)$, and evaluates to $-1$. Therefore we need to add a scalar with twist $\frac34\(\frac{c-1}{24}\)$ with degeneracy $1$. When we look at $n=4, 6, \ldots$, we only need to consider odd spins.

For $j\equiv1~(\text{mod}~2)$, 
\be
\frac{S(j, 0;2) - S(j, 1;2)}2 = -1
\ee
which means we also need to add a single state at $\widetilde T_2$.

\subsection*{$\mathbf{n=3}$}

Since $3$ is prime the only term we need to worry about is the term at twist $0$, aka the vacuum. The term
\be
\frac{S(j, 0;3) - S(j, -1; 3)}3 
\ee
is most negative for $j\equiv1~(\text{mod}~3)$, and evaluates to $-1$. Therefore we need to add a scalar with twist $\frac89\(\frac{c-1}{24}\)$ with degeneracy $1$. When we look at $n=6, 9, \ldots$, we need to only consider spins $1~(\text{mod}~3)$. 
At $j\equiv1~(\text{mod}~3)$,
\be
\frac{S(j, 0;3) - S(j, 1; 3)}3 = 0,
\ee
so no states at reduced twist $\widetilde T_3$ are necessary. 

\subsection*{$\mathbf{n=4}$}

This is the first nontrivial example. We need to worry about the vacuum, and the one state we added at twist $\frac{c-1}{32}$. From the vacuum, we learned that the only states that have a chance of being negative at high spin are the ones that have odd spin (and also spins that are $1~(\text{mod}~3)$ but that is irrelevant since 3 and 4 are coprime). 

From the vacuum we need to consider 
\be
\frac{S(j, 0;4) - S(j, -1;4)}4
\ee
For $j\equiv1~(\text{mod}~4)$, this is $-\frac12$ and for $j\equiv3~(\text{mod}~4)$ this is $\frac12$.
From the twist $\frac{c-1}{32}$ state we need to consider
\be
\frac{S(j,0;2)}2
\ee
For odd $j$ this is $-\frac12$. Therefore the states that are $1~(\text{mod}~4)$ currently are negative; we need to add a state with twist $\frac{15}{16}\(\frac{c-1}{24}\)$ with degeneracy $\frac12 + \frac12 = 1$ to cancel this. When we look at future $s=8, 12, 16, \ldots$, we need to only consider $1~(\text{mod}~4)$ for negativity.
For states $j\equiv1~(\text{mod}~4)$,
\be
\frac{S(j, 0;4) - S(j, 1;4)}4 + \frac{S(j,0;2)}2 = 0,
\ee
so no states with reduced twist $\widetilde T_4$ are necessary.

\subsection*{$\mathbf{n=5}$}
The term
\be
\frac{S(j, 0;5) - S(j, -1; 5)}5 
\ee
is most negative for $j\equiv1~(\text{mod}~5)$, where it evaluates to $-\frac{5+\sqrt{5}}{10} \sim -0.7236$. We therefore need to add $\frac{5+\sqrt{5}}{10}$ states with reduced twist $T_5$. For $j\equiv1~(\text{mod}~5)$, 
\be
\frac{S(j, 0;5) - S(j, 1; 5)}5 = -\frac{5-\sqrt{5}}{10} \sim -0.2764.
\ee
We therefore also need to add $\frac{5-\sqrt{5}}{10}$ states at reduced twist $\widetilde T_5$. 

\subsection*{$\mathbf{n=6}$}

From the $n=2,3$ calculations, we only need to consider states of spin $j\equiv1~(\text{mod}~6)$. The term $T_2$ contributes $\frac{S(j,0;3)}3$ and the term $T_3$ contributes $\frac{S(j,0;2)}2$ to the density of states with order $\frac{\exp{\(\frac{4\pi}{6}\sqrt{(\frac{c-1}{24})j}\)}}{\sqrt{j\(\bar h -\frac{c-1}{24}\)}}$. If $j\equiv1~(\text{mod}~6)$ then
\be
\frac{S(j,0;6) - S(j,-1;6)}{6} + \frac{S(j,0;3)}3 + \frac{S(j,0;2)}2 = -1.
\ee
Thus we add $1$ state at reduced twist $T_6$. Similarly the state $\widetilde T_2$ contributes $\frac{S(j,0;3)}3$ to the density of states with order $\frac{\exp{\(\frac{4\pi}{6}\sqrt{(\frac{c-25}{24})j}\)}}{\sqrt{j\(\bar h -\frac{c-1}{24}\)}}$. If $j\equiv1~(\text{mod}~6)$ then
\be
\frac{S(j,0;6) - S(j,1;6)}{6} + \frac{S(j,0;3)}3  = 0
\ee
so no states are needed with reduced $\widetilde T_6$.

\subsection*{$\mathbf{n=7}$}

The term
\be
\frac{S(j, 0;7) - S(j, -1; 7)}7 
\ee
is most negative for $j\equiv2~(\text{mod}~7)$, where it evaluates to $-\frac{3+4\sin{\(\frac{3\pi}{14}\)}}{7} \sim -0.7849$. Thus we need to add $\frac{3+4\sin{\(\frac{3\pi}{14}\)}}{7}$ states with reduced twist $T_7$. 

However, if $j\equiv2~(\text{mod}~7)$, then 
\be
\frac{S(j, 0;7) - S(j, 1; 7)}7 = \frac{-1+4\cos\(\frac{\pi}{7}\)-2\sin\(\frac{3\pi}{14}\)}{7} \sim 0.1938 > 0.
\ee
Thus we do not need to add any states at reduced $\widetilde T_7$. 

So far our analysis shows that the addition of the states above renders the spectrum positive for all large spins $j$. For sufficiently large central charge, this moreover is sufficient to render the spectrum positive for all finite spins. If the central charge is large enough, the negativity coming from the $s=8, 9, \ldots$ terms will be suppressed; since $\frac{3+4\sin{\(\frac{3\pi}{14}\)}}{7} > -\frac{2S(j,0;7)-S(j,1;7)-S(j,-1;7)}{7}$ for all $j$, the term added is sufficient to render the density of states at all spins mod $7$ positive. 

In practice we find numerically that for central charge $c>c_*$, the spectrum is positive everywhere, with
\be
c_* \sim 3237.7.
\ee

\section{Full orbifold sum}
\label{app:fullorbifoldspectrum}

In this appendix, we write the full, finite partition function that includes a sum over all orbifolds $T_N$ and $\widetilde T_N$. As discussed in section \ref{sec:discuss}, the sum over all states with reduced twist $T_N, \widetilde T_N$ results in a convergent, finite, modular invariant partition function. The sum over $T_N$ becomes
\begin{align}
\sum_{N=1}^\infty Z_{T_N}(\tau,\bar\tau) =& \frac1{\eta(q)\eta(\bar q)} \Bigg(\sum_{m=1}^\infty{(4\pi\xi\tau_2)^m\over m!}\zeta(2m) +  \sum_{m=1}^{\infty} \frac{\Gamma(m)(16\pi \xi)^m\zeta(2m)^2}{\zeta(2m+1)(2m)!\tau_2^m}  \nonumber\\&+\sum_{j=1}^{\infty} 2\(e^{2\pi i j\tau_1} + e^{-2\pi i j \tau_1}\)\sum_{m=1}^\infty  \frac{(16\pi^2 \xi)^m \sigma_{2m}(j)\zeta(2m)}{(2m)! j^m \zeta(2m+1)} K_m(2\pi j\tau_2)\Bigg),
\label{eq:sumOverOrbifoldsApp}
\end{align}
with $\xi = \frac{c-1}{24}$, while the sum over $\widetilde T_N$ becomes the same equation but with $\xi \to \xi-1 = \frac{c-25}{24}$. The $N=1$ terms in these expressions also correspond to two of the four terms in the vacuum character. Finally, the remaining two terms are given by a $PSL(2,\mathbb Z)$ sum of a seed with spin $J=\pm 1$, and reduced twist $T=-\frac{c-1}{24}$. This sum is given in \cite{Keller:2014xba}, which we rewrite in terms of hypergeometric functions as
\begin{align}
&\frac{1}{\eta(q)\eta(\bar{q})}\Bigg[2e^{2\pi \tau_2 E}\cos(2\pi \tau_1) + 4 + \sum_{m=1}^{\infty} \frac{4\pi^{m+\frac12}T_m(E) {\tau_2}^{-m}}{m\Gamma(m+\frac12) \zeta(2m+1)} \nonumber\\
&+ \sum_{m=1}^\infty \sum_{j=1}^{\infty} \frac{4\sqrt \pi 2^m \pi^{2m} j^m \cos\(2\pi j \tau_1\)}{\Gamma(m+\frac12)}\Bigg(\mathcal{Z}_{j,1}(m+\frac12) \sum_{a=0}^m (-1)^{a+m} \frac{_1F_1(-a,1-2a+m,-4\pi \tau_2 j E)}{a!(m-2a)!}\frac{K_a(2\pi \tau_2 j)}{(4\pi \tau_2 j)^a} \nonumber \\  &~~~~~~~~~~~~~~~~~~~~~~~~~~~~~~~~~~~~~~~~~~~~~+\mathcal{Z}_{-j,1}(m+\frac12) \sum_{a=0}^m (-1)^a \frac{_1F_1(-a,1-2a+m,4\pi \tau_2 j E)}{a!(m-2a)!}\frac{K_a(2\pi \tau_2 j)}{(4\pi \tau_2 j)^a}\Bigg) \nonumber\\
&+  \sum_{j=1}^{\infty} 4\cos\(2\pi j \tau_1\) \(\mathcal{Z}_{j,1}(\frac12) + \mathcal{Z}_{-j,1}(\frac12)\) K_0(2\pi \tau_2 j) \Bigg],
\label{eq:nonullsub}
\end{align} 
where $T_m$ is a Chebyshev polynomial. In (\ref{eq:nonullsub}), we set
\be
E = \frac{c-13}{12}
\ee and the function $\mathcal{Z}_{j,J}(m)$ is a Kloosterman zeta function defined as
\be
\mathcal{Z}_{j,J}(m) = \sum_{c=1}^{\infty} c^{-2m} S(j,J;c).
\label{eq:Kloos}
\ee
The expression (\ref{eq:Kloos}) is convergent for every line in (\ref{eq:nonullsub}) except the last; there it has to be defined via analytic continuation. For convenience we can rewrite (\ref{eq:nonullsub}) in terms of a generalization of the non-holomorphic Eisenstein series:
\be
\frac{1}{\sqrt{\tau_2}\eta(q)\eta(\bar q)}\sum_{m=0}^\infty \frac{(2\pi E)^m}{m!} \hat E\(m+\frac12, \tau, \bar\tau\)
\ee
where
\begin{align}
\hat E\(m+\frac12, \tau,\bar\tau\) &= 2\cos\(2\pi \tau_1\) \tau_2^{m+\frac12} + \frac{2\sqrt\pi\Gamma(m)}{\Gamma(m+\frac12)\zeta(2m+1)} \tau_2^{\frac12-m}\nonumber\\
&+\sum_{a=1}^{\infty} \frac{2(-1)^{a}\pi^{2a+\frac12}\Gamma(a+m)}{\Gamma(a+1)\Gamma(2a+m+\frac12)\zeta(4a+2m+1)}\tau_2^{\frac12 - m-2a} \nonumber\\
&+ \sum_{j=1}^{\infty} \sum_{a =0}^{\infty} \sum_{b=0}^{\left \lfloor{a/2}\right \rfloor} \frac{4(-1)^{b} 2^{a-2b} j^{a-b+m} \pi^{\frac12+2a-b+m}\cos\(2\pi j \tau_1\) }{\Gamma(a-2b+1) \Gamma\(a+ m + \frac12\) \Gamma(b + 1)} \tau_2^{\frac12 -b} K_{b+m}\(2\pi j \tau_2\)  \nonumber\\
&~~~~~~~~~~~~~~~~~~~~~~\times \left[(-1)^{a}\mathcal{Z}_{j,1}(m + a+ \frac12) +\mathcal{Z}_{j,-1}(m + a+ \frac12)\right].
\label{eq:omgawesome}
\end{align}
Each of the $\hat E$ functions in (\ref{eq:omgawesome}) are the $PSL(2,\mathbb Z)$ sum of $2 {\tau_2}^{m+\frac12} \cos(2\pi \tau_1)$ and are by construction modular invariant. It would be straightforward to generalize this to a $PSL(2,\mathbb Z)$ sum of ${\tau_2}^{m+\frac12} e^{2\pi i J \tau_1}$ for any integer $J$, but we will do not so here.

The final partition function is the sum of (\ref{eq:sumOverOrbifoldsApp}) and (\ref{eq:sumOverOrbifoldsApp}) with $\xi$ replaced with $\xi-1$, minus (\ref{eq:nonullsub}). 

\bibliographystyle{JHEP}
\bibliography{pureGravity}

\providecommand{\href}[2]{#2}\begingroup\raggedright\begin{thebibliography}{10}

\bibitem{Banados:1992wn}
M.~Banados, C.~Teitelboim, and J.~Zanelli, {\it {The Black hole in
  three-dimensional space-time}},  {\em Phys. Rev. Lett.} {\bf 69} (1992)
  1849--1851, [\href{http://arxiv.org/abs/hep-th/9204099}{{\tt
  hep-th/9204099}}].

\bibitem{Brown:1986nw}
J.~D. Brown and M.~Henneaux, {\it {Central Charges in the Canonical Realization
  of Asymptotic Symmetries: An Example from Three-Dimensional Gravity}},  {\em
  Commun. Math. Phys.} {\bf 104} (1986) 207--226.

\bibitem{Witten:2007kt}
E.~Witten, {\it {Three-Dimensional Gravity Revisited}},
  \href{http://arxiv.org/abs/0706.3359}{{\tt arXiv:0706.3359}}.

\bibitem{Heemskerk:2009pn}
I.~Heemskerk, J.~Penedones, J.~Polchinski, and J.~Sully, {\it {Holography from
  Conformal Field Theory}},  {\em JHEP} {\bf 10} (2009) 079,
  [\href{http://arxiv.org/abs/0907.0151}{{\tt arXiv:0907.0151}}].

\bibitem{Cardy:1986ie}
J.~L. Cardy, {\it {Operator Content of Two-Dimensional Conformally Invariant
  Theories}},  {\em Nucl. Phys.} {\bf B270} (1986) 186--204.

\bibitem{Strominger:1996sh}
A.~Strominger and C.~Vafa, {\it {Microscopic origin of the Bekenstein-Hawking
  entropy}},  {\em Phys. Lett.} {\bf B379} (1996) 99--104,
  [\href{http://arxiv.org/abs/hep-th/9601029}{{\tt hep-th/9601029}}].

\bibitem{Strominger:1997eq}
A.~Strominger, {\it {Black hole entropy from near horizon microstates}},  {\em
  JHEP} {\bf 02} (1998) 009, [\href{http://arxiv.org/abs/hep-th/9712251}{{\tt
  hep-th/9712251}}].

\bibitem{Hartman:2014oaa}
T.~Hartman, C.~A. Keller, and B.~Stoica, {\it {Universal Spectrum of 2d
  Conformal Field Theory in the Large c Limit}},  {\em JHEP} {\bf 09} (2014)
  118, [\href{http://arxiv.org/abs/1405.5137}{{\tt arXiv:1405.5137}}].

\bibitem{Hellerman:2009bu}
S.~Hellerman, {\it {A Universal Inequality for CFT and Quantum Gravity}},  {\em
  JHEP} {\bf 08} (2011) 130, [\href{http://arxiv.org/abs/0902.2790}{{\tt
  arXiv:0902.2790}}].

\bibitem{Friedan:2013cba}
D.~Friedan and C.~A. Keller, {\it {Constraints on 2d CFT partition functions}},
   {\em JHEP} {\bf 10} (2013) 180, [\href{http://arxiv.org/abs/1307.6562}{{\tt
  arXiv:1307.6562}}].

\bibitem{Collier:2016cls}
S.~Collier, Y.-H. Lin, and X.~Yin, {\it {Modular Bootstrap Revisited}},  {\em
  JHEP} {\bf 09} (2018) 061, [\href{http://arxiv.org/abs/1608.06241}{{\tt
  arXiv:1608.06241}}].

\bibitem{Afkhami-Jeddi:2019zci}
N.~Afkhami-Jeddi, T.~Hartman, and A.~Tajdini, {\it {Fast Conformal Bootstrap
  and Constraints on 3d Gravity}},  {\em JHEP} {\bf 05} (2019) 087,
  [\href{http://arxiv.org/abs/1903.06272}{{\tt arXiv:1903.06272}}].

\bibitem{Hartman:2019pcd}
T.~Hartman, D.~Mazac, and L.~Rastelli, {\it {Sphere Packing and Quantum
  Gravity}},  {\em JHEP} {\bf 12} (2019) 048,
  [\href{http://arxiv.org/abs/1905.01319}{{\tt arXiv:1905.01319}}].

\bibitem{Maloney:2007ud}
A.~Maloney and E.~Witten, {\it {Quantum Gravity Partition Functions in Three
  Dimensions}},  {\em JHEP} {\bf 02} (2010) 029,
  [\href{http://arxiv.org/abs/0712.0155}{{\tt arXiv:0712.0155}}].

\bibitem{Keller:2014xba}
C.~A. Keller and A.~Maloney, {\it {Poincare Series, 3D Gravity and CFT
  Spectroscopy}},  {\em JHEP} {\bf 02} (2015) 080,
  [\href{http://arxiv.org/abs/1407.6008}{{\tt arXiv:1407.6008}}].

\bibitem{Benjamin:2019stq}
N.~Benjamin, H.~Ooguri, S.-H. Shao, and Y.~Wang, {\it {Light-cone modular
  bootstrap and pure gravity}},  {\em Phys. Rev.} {\bf D100} (2019), no.~6
  066029, [\href{http://arxiv.org/abs/1906.04184}{{\tt arXiv:1906.04184}}].

\bibitem{Maldacena:2000hw}
J.~M. Maldacena and H.~Ooguri, {\it {Strings in AdS(3) and SL(2,R) WZW model
  1.: The Spectrum}},  {\em J. Math. Phys.} {\bf 42} (2001) 2929--2960,
  [\href{http://arxiv.org/abs/hep-th/0001053}{{\tt hep-th/0001053}}].

\bibitem{Eberhardt:2019qcl}
L.~Eberhardt and M.~R. Gaberdiel, {\it {String theory on AdS$_3$ and the
  symmetric orbifold of Liouville theory}},  {\em Nucl. Phys.} {\bf B948}
  (2019) 114774, [\href{http://arxiv.org/abs/1903.00421}{{\tt
  arXiv:1903.00421}}].

\bibitem{Krasnov:2000ia}
K.~Krasnov, {\it {3-D gravity, point particles and Liouville theory}},  {\em
  Class. Quant. Grav.} {\bf 18} (2001) 1291--1304,
  [\href{http://arxiv.org/abs/hep-th/0008253}{{\tt hep-th/0008253}}].

\bibitem{Krasnov:2001ui}
K.~Krasnov, {\it {Lambda less than 0 quantum gravity in 2+1 dimensions. 1.
  Quantum states and stringy S matrix}},  {\em Class. Quant. Grav.} {\bf 19}
  (2002) 3977--3998, [\href{http://arxiv.org/abs/hep-th/0112164}{{\tt
  hep-th/0112164}}].

\bibitem{Krasnov:2002rn}
K.~Krasnov, {\it {Lambda less than 0 quantum gravity in (2+1)-dimensions. 2.
  Black hole creation by point particles}},  {\em Class. Quant. Grav.} {\bf 19}
  (2002) 3999--4028, [\href{http://arxiv.org/abs/hep-th/0202117}{{\tt
  hep-th/0202117}}].

\bibitem{Saad:2019lba}
P.~Saad, S.~H. Shenker, and D.~Stanford, {\it {JT gravity as a matrix
  integral}},  \href{http://arxiv.org/abs/1903.11115}{{\tt arXiv:1903.11115}}.

\bibitem{Stanford:2019vob}
D.~Stanford and E.~Witten, {\it {JT Gravity and the Ensembles of Random Matrix
  Theory}},  \href{http://arxiv.org/abs/1907.03363}{{\tt arXiv:1907.03363}}.

\bibitem{Alday:2019vdr}
L.~F. Alday and J.-B. Bae, {\it {Rademacher Expansions and the Spectrum of 2d
  CFT}},  \href{http://arxiv.org/abs/2001.00022}{{\tt arXiv:2001.00022}}.

\bibitem{Maxfield:2019hdt}
H.~Maxfield, {\it {Quantum corrections to the BTZ black hole extremality bound
  from the conformal bootstrap}},  {\em JHEP} {\bf 12} (2019) 003,
  [\href{http://arxiv.org/abs/1906.04416}{{\tt arXiv:1906.04416}}].

\bibitem{Maldacena:1998bw}
J.~M. Maldacena and A.~Strominger, {\it {AdS(3) black holes and a stringy
  exclusion principle}},  {\em JHEP} {\bf 12} (1998) 005,
  [\href{http://arxiv.org/abs/hep-th/9804085}{{\tt hep-th/9804085}}].

\bibitem{Dijkgraaf:2000fq}
R.~Dijkgraaf, J.~M. Maldacena, G.~W. Moore, and E.~P. Verlinde, {\it {A Black
  hole Farey tail}},  \href{http://arxiv.org/abs/hep-th/0005003}{{\tt
  hep-th/0005003}}.

\bibitem{Deser:1983tn}
S.~Deser, R.~Jackiw, and G.~'t~Hooft, {\it {Three-Dimensional Einstein Gravity:
  Dynamics of Flat Space}},  {\em Annals Phys.} {\bf 152} (1984) 220.

\bibitem{Collier:2018exn}
S.~Collier, Y.~Gobeil, H.~Maxfield, and E.~Perlmutter, {\it {Quantum Regge
  Trajectories and the Virasoro Analytic Bootstrap}},  {\em JHEP} {\bf 05}
  (2019) 212, [\href{http://arxiv.org/abs/1811.05710}{{\tt arXiv:1811.05710}}].

\bibitem{Kusuki:2018wpa}
Y.~Kusuki, {\it {Light Cone Bootstrap in General 2D CFTs and Entanglement from
  Light Cone Singularity}},  {\em JHEP} {\bf 01} (2019) 025,
  [\href{http://arxiv.org/abs/1810.01335}{{\tt arXiv:1810.01335}}].

\bibitem{Dabholkar:2014ema}
A.~Dabholkar, J.~Gomes, and S.~Murthy, {\it {Nonperturbative black hole entropy
  and Kloosterman sums}},  {\em JHEP} {\bf 03} (2015) 074,
  [\href{http://arxiv.org/abs/1404.0033}{{\tt arXiv:1404.0033}}].

\bibitem{Murthy:2009dq}
S.~Murthy and B.~Pioline, {\it {A Farey tale for N=4 dyons}},  {\em JHEP} {\bf
  09} (2009) 022, [\href{http://arxiv.org/abs/0904.4253}{{\tt
  arXiv:0904.4253}}].

\bibitem{Banerjee:2008ky}
N.~Banerjee, D.~P. Jatkar, and A.~Sen, {\it {Asymptotic Expansion of the N=4
  Dyon Degeneracy}},  {\em JHEP} {\bf 05} (2009) 121,
  [\href{http://arxiv.org/abs/0810.3472}{{\tt arXiv:0810.3472}}].

\bibitem{Giombi:2008vd}
S.~Giombi, A.~Maloney, and X.~Yin, {\it {One-loop Partition Functions of 3D
  Gravity}},  {\em JHEP} {\bf 08} (2008) 007,
  [\href{http://arxiv.org/abs/0804.1773}{{\tt arXiv:0804.1773}}].

\bibitem{Cotler:2018zff}
J.~Cotler and K.~Jensen, {\it {A theory of reparameterizations for AdS$_3$
  gravity}},  {\em JHEP} {\bf 02} (2019) 079,
  [\href{http://arxiv.org/abs/1808.03263}{{\tt arXiv:1808.03263}}].

\bibitem{Oblak:2016eij}
B.~Oblak, {\em {BMS Particles in Three Dimensions}}.
\newblock PhD thesis, Brussels U., 2016.
\newblock \href{http://arxiv.org/abs/1610.08526}{{\tt arXiv:1610.08526}}.

\bibitem{Collier:2019weq}
S.~Collier, A.~Maloney, H.~Maxfield, and I.~Tsiares, {\it {Universal Dynamics
  of Heavy Operators in CFT$_2$}},  \href{http://arxiv.org/abs/1912.00222}{{\tt
  arXiv:1912.00222}}.

\end{thebibliography}\endgroup
\end{document}